\documentstyle[twoside,12pt]{article}
\input{epsf.sty}
\newcommand{\ooo}{\!\not \! o}

\makeatletter
\@addtoreset{equation}{section}
\makeatother

\def\a{\alpha}
\def\b{\beta}
\def\c{\chi}

\def\pr{\prime}
\def\d{\delta}
\def\e{\epsilon}   
\def\g{\gamma}

\def\k{\kappa}     
\def\l{\lambda}
\def\m{\mu}

\def\n{\nu}

\def\p{\pi}       

\def\x{\xi}
\def\z{\zeta}
\def\D{\Delta}

\def\L{\Lambda}

\def\cD{{\cal D}}

\newcommand{\del}{\partial}
\newcommand{\dpp}{\partial^{\prime}}
\newcommand{\av}[1]{\mbox{$\langle #1 \rangle$}}

\newcommand{\half}{\mbox{{\normalsize $\frac{1}{2}$}} }
\newcommand{\quart}{\mbox{{\small $\frac{1}{4}$}} }

\newcommand{\third}{\mbox{{\small $\frac{1}{3}$}} }

\newcommand{\ra}{\rightarrow}

\newcommand{\apgt}{ \mbox{}_{\textstyle \sim}^{\textstyle > }     }

\newcommand{\bt}{\beta}

\newcommand{\nnn}{\nonumber \\}

\newcommand{\mh}{\hat{\mu}}

\newcommand{\be}{\begin{equation}}
\newcommand{\ee}{\end{equation}}
\newcommand{\bea}{\begin{eqnarray}}
\newcommand{\eea}{\end{eqnarray}}
\newcommand{\eq}{\ref}
\newcommand{\beq}{\begin{equation}}
\newcommand{\eeq}{\end{equation}}
\newcommand{\cc}{\cite}
\newcommand{\lb}{\label}
\newcommand{\lsim}{\stackrel{<}{\sim}} 

\def \3{\ss}

\def\footnoteitem(#1)#2{
\begin{list}{#1}{\labelwidth4.0mm \leftmargin7.0mm
\labelsep2.5mm \rightmargin7.0mm \parsep0.5ex plus0.2ex minus0.1ex
\itemsep0ex plus0.2ex }
\item #2
\end{list}
}
\begin{document}

\headsep=0.0cm
\vsize 25.0truecm
\topmargin=0cm
\topskip=0cm

\begin{titlepage}

\rightline{\bf hep-lat/9406018}
\rightline{UCSD/PTH 94-08}
\vskip 3mm
\rightline{June 1994}

\baselineskip=20pt plus 1pt
\vskip 0.5cm

\centerline{\LARGE Failure of the Regge approach in two}
\centerline{\LARGE dimensional quantum gravity}
\vskip 2.7cm

\centerline{
{\bf Wolfgang Bock
}~~~~~~~\raisebox{-1.0ex}{and}~~~~~~~{\bf Jeroen C. Vink}}
\centerline{
{\sf \hspace{-.01cm} bock@sdphjk.ucsd.edu}~~~~~~~~~~~~~~~{\sf
vink@yukawa.ucsd.edu}}
\medskip
\centerline{\it University of California, San Diego}
\centerline{\it Department of Physics-0319}
\centerline{\it 9500 Gilman Drive, La Jolla, CA 92093-0319}
\centerline{\it USA}

\vskip 2.7cm
\baselineskip=12pt plus 1pt
\parindent 20pt
\centerline{\bf Abstract}
\textwidth=6.0truecm
\medskip

\frenchspacing
Regge's method for regularizing euclidean quantum gravity is applied to
two dimensional gravity.  We use two  different strategies to simulate
the Regge  path integral at a fixed value of the total area: A standard
Metropolis simulation combined with a histogramming method and a direct
simulation using a Hybrid Monte Carlo algorithm. Using topologies with
genus zero and two and a scale  invariant integration measure, we show
that the Regge method does not reproduce the value  of the   string
susceptibility of the continuum model.  We show that  the string
susceptibility  depends  strongly on  the choice of the measure in
the path integral. We argue that the failure of the Regge approach
is due to spurious contributions of reparametrization degrees of
freedom to the path integral.
\nonfrenchspacing

\end{titlepage}

\textheight=20.5cm
\headsep=2.0cm
\vsize 21truecm
%
%
\section{Introduction}
It is still an unsolved problem how a quantum theory of
gravitation can be formulated using an euclidean path integral.
Well-known problems are the unboundedness of the euclidean action,
the perturbative non-renormalizability of the Einstein-Hilbert
action and finding a consistent regularization, for instance by
discretizing space-time, such that the path integral can be
studied non-perturbatively with numerical methods.
Here we shall focus on the problem of finding a regularization of the
euclidean path integral.

Most studies so far have either used
dynamical triangulation  \cc{DT,Da85} or Regge calculus \cc{Regge}.
For recent review articles we refer the
reader to refs.~\cc{Ka92,REVIEW}.
For an approach along a different line,
see refs. \cc{Sm79,Me90}.
In both cases the path integral is replaced by a summation over
simplicial manifolds. These simplicial
manifolds are built up from $d$ dimensional simplices,
triangles in $d=2$ dimensions, which are glued together along their $d-1$
dimensional subsimplices, the edges in two dimensions.
In the dynamical triangulation approach one keeps the edge lengths
fixed and replaces the integration in the path integral by a summation
over simplicial structures with all possible connectivities and numbers
of simplices.
In the Regge method the edge lengths are the dynamical degrees of freedom,
while the connectivities are usually kept fixed.
Here the path integral is replaced by an integration over the edge lengths.
Both formulations have appealing features, but  are also subject to
criticism.

In both approaches the metric tensor field, which is the integration
variable in the continuum path integral, is replaced by a simplicial
structure.
With dynamical triangulation, the Einstein-Hilbert action is automatically
bounded from below; in the Regge approach one usually adds a curvature
square term to make the action bounded \cc{HaWi84}. It is far from
clear, however, how the continuum integration measure, which includes
gauge fixing and Faddeev-Popov terms, has to be translated to a
summation  over these simplicial structures.

With dynamical triangulation a natural choice is to simply sum over
all triangulated surfaces, only weighted by the action. Since it is
impossible to use dynamical triangulation to study small fluctuations
around a flat space in perturbation theory, the only way to test
this conjecture is through numerical simulations. This should reveal
the existence of a second order phase transition at which
renormalized quantities, such as the volume and curvature fluctuations,
become independent of the details of the underlying discrete structure.
Studies of the phase structure have given some evidence for the existence
of such a critical point in four dimensions
\cc{PTfour,AmJu94}, but the situation is far from clear
\cc{CaKo94,BaSm94}. It has furthermore  been  questioned if
the  Monte Carlo methods which are presently
in use are  ergodic \cc{NaBe93}.
The strongest support for the dynamical triangulation approach comes
from its success in two dimensional gravity. Here it has been shown in
detail, that the results of the continuum theory are reproduced
\cc{Ka92,REVIEW,AmJa93}.

In the Regge approach it is less easy to choose a natural measure for
the integration over the link lengths. In contrast to the dynamical
triangulation approach, it is possible to study the Regge path integral
in perturbation theory. Here it has been shown that the path integral
with a simple (scale invariant) measure, corresponds to the continuum
path integral without gauge fixing or Faddeev-Popov terms
\cc{RoWi81}.
This suggests that also outside perturbation theory, the integration
includes degrees of freedom which correspond to coordinate
transformations; we shall discuss this in more detail below.
To remove these spurious degrees of freedom, one would expect that the
measure should include gauge fixing and Faddeev-Popov
terms. One has often commented on this issue, but  in actual simulations
only  the scale invariant measure or
simple variations thereof have been used.

Even though the justification of the measure used in
dynamical triangulation appears to be equally obscure as
in the Regge approach, this latter approach has been lacking
the compelling success of the dynamical triangulation method in two
dimensional gravity.
This has motivated some time ago an investigation of the Regge approach in two
dimensional gravity \cc{GrHa91}. This numerical study claimed
that also the Regge method reproduces
correctly the scaling of the partition function, in particular the so-called
string susceptibility of two dimensional continuum gravity with central
charge zero was found. The string susceptibility
was  not found to coincide  with the continuum prediction
for non-zero central      charge.
In the present work we describe the results of a similar study of the
central charge zero model, which
we believe improves several of the shortcomings of ref. \cc{GrHa91}.
We shall use the same Regge action,
which includes an $R^2$ term, and
the same scale invariant integration measure as
in ref.~\cc{GrHa91}. Unfortunately, we find wrong values for the
string susceptibility, when the simplicial structure has the topology of
a sphere or a ``bi-torus'' (a manifold with two handles).
We will show  explicitly that the scaling
of the Regge partition function depends very sensitively on the details
of the integration measure.

The paper is organized as follows. In sect.~2 we review the continuum
scaling relations for the path integral over surfaces with a fixed
area. We discuss in some detail how the Regge skeletons can
be viewed as discretizations of the metric field of continuum manifolds.
Sect. 3 deals with the derivation of  scaling
relations for the Regge path integral over skeletons with
a fixed area. We shall show that  these  scaling relations reveal
a sensitive dependence of the string susceptibility on
the integration measure. In sect.~4 we discuss the technical aspects
of our simulation methods.
As a novelty, a Hybrid Monte Carlo algorithm is introduced which can
be used to simulate the fixed area Regge path integral. Our numerical
results are presented in sect.~5 and
a brief summary and some final remarks are contained in sect.~6.
\section{Regge approach to two dimensional gravity}
\subsection{Continuum scaling relations}
Let us consider the euclidean path integral
for two dimensional pure gravity (the central charge zero model),
\bea
Z(A) & = & \int \frac{\cD g}{{\rm Vol(Diff)}} \; e^{-S(g)} \;
       \d(\int d^2x \sqrt{g} - A) \;, \lb{ZZPART} \\
 S(g)& = &\int d^2x \; \sqrt{g} \; (\l + \k R + \b R^2)  \;.
                 \label{SCONT}
\eea
In two dimensions the curvature term $\int d^2 x \sqrt{g}R=4\p\c$
is a topological invariant (Gauss-Bonnet theorem)
depending only on the Euler characteristic
$\c=2(1-h)$, where $h$ denotes the number of handles or genus of the surface.
The genus is zero for the sphere, one for the torus and two
for the bi-torus.
We have included a $R^2$ term in the action with a coupling constant
$\b$ which has a dimension of 1/mass$^2$. The coupling constant $\k$ is
dimensionless and
the cosmological constant $\l$ has a dimension of mass$^2$.
The delta function in (\eq{ZZPART})
imposes the constraint that the area of the surface is equal to $A$.
Formally the measure is normalized by the infinite volume of the
diffeomorphism group, which should cancel the infinite factor that
arises from the integration over the gauge (reparametrization) degrees
of freedom in the path integral. The continuum measure
$\cD g$ is ill-defined because it
involves an infinite product over all space-time points.

On a formal level, i.e. ignoring the measure, one would
expect that
\be
Z(A) \propto A^{-1} \exp [ -\l A]\;, \lb{NSCA}
\ee
if we set $\b$ in the action (\eq{SCONT}) equal to zero. However, after
regularization  and fixing the gauge using the usual  Faddeev-Popov
procedure, one finds that the measure breaks the conformal (scale)
symmetry and the $A$ dependence of the partition function is given by
\be
 Z(A)\propto A^{\g_{{\rm str}-3}} \exp [-\l_R A] \;,  \label{ZCONT}
\ee
where $\l_R$ is the renormalized cosmological constant.
The string susceptibility $\g_{{\rm str}}$ depends on the genus $h$
of the surface,
\be
\g_{{\rm str}} =  2 - \frac{5}{2} \; (1-h)  \;. \lb{STRING}
\ee
The relations (\eq{ZCONT}) and (\eq{STRING})
were first derived for the special case $h=0$
using the light cone gauge \cc{KnPo88}
and were later confirmed \cc{Da88} and generalized to
topologies of arbitrary genus $h$ \cc{DiKa89} using the conformal gauge.
Quantum fluctuations are seen to lead to a deviation from
the naive scaling dimension of the partition function for $h\ne1$
and renormalize the cosmological constant.

For small, but non-zero $\b$ one expects that the result (\ref{ZCONT})
remains valid in the regime where $A \gg 1/\b$, because the $R^2$ term
should only modify the short distance behavior of the theory. The $R^2$
model for $A/\b\ra 0$ has been studied in ref. \cc{KaNa93}.
Using the dynamical triangulation method it has been demonstrated
numerically that the addition of a small
$R^2$ term did  not affect the value of the string
susceptibility \cc{Da85,BoKa86}.

In order to study this two dimensional gravity model using the Regge
regularization, one first has to express
the action (\eq{SCONT}) in terms of the edge lengths and then
integrate over all length variables with an appropriate choice for
the measure.
In order to expose more clearly the role of reparametrization
invariance  we shall recall briefly  that the Regge action can be viewed
as a discretization of the continuum Einstein-Hilbert action.
This  establishes a link between the edge lengths of
a Regge skeleton and the metric field in the continuum and furthermore
sheds light on the problem of choosing the integration measure in the Regge
approach.
\subsection{Regge discretization}
Usually the Regge skeletons are considered as special, piecewise flat,
continuum manifolds. Here we want to take a somewhat different point of
view, in which a Regge skeleton is a discretized version of an
underlying continuum manifold which is defined by a metric tensor
field $g_{\m\n}(x)$.
For simplicity we shall assume that the continuum
manifold $T$ has the topology of a torus.
Then we can choose global coordinates $\{x=(x_1,x_2)\}$ on this
surface with metric $g_{\m \n}(x)$.
The coordinates are a (globally defined) map from a flat unit torus
$T_0=\{x|x_1\in [0,1), x_2\in[0,1)\}$, into the curved surface of $T$.
We can construct a grid on the curved torus $T$  by  first defining
the  grid $\L=\{x| x_\m=n_\m a_0; n_\m=1,\cdots,N; a_0=1/N\}$ on $T_0$,
as shown in fig.~\ref{fig1} and then
mapping it onto
the curved surface. In this way we can obtain triangulations of $T$
where the triangles are  spanned by the triplets of coordinates
$\{(n_1 a, n_2 a)$, $((n_1+1) a ,n_2 a), ((n_1+1) a,(n_2+1) a)\}$ and
$\{(n_1 a ,n_2 a ),(n_1 a ,(n_2+1) a),((n_1+1) a,(n_2+1) a)\}$,
cf. fig.~\ref{fig1}.
\begin{figure} [tttttbbb]
\vspace{-1.6cm}
 \centerline{ \epsfysize=12cm \epsfbox{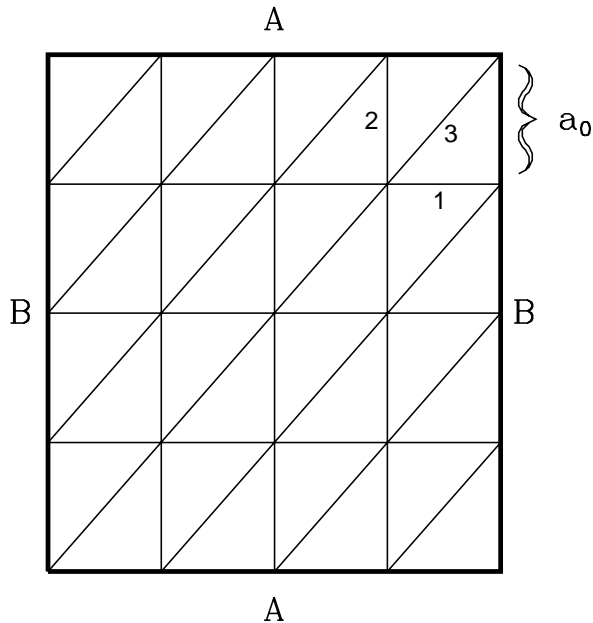} }
\vspace{-2.5cm}
\caption{{\em Triangulation  of the torus with $N=4$.
              The boundaries with
               the same labels have to be
              identified. The lattice spacing is denoted
              by $a_0$. Each vertex in the skeleton
              has coordination number six.}}
\lb{fig1}
\end{figure}
The edge lengths to be used in the Regge skeleton,
can be defined  in various ways: one can take
the proper length of images of the
corresponding edges on the unit torus $T_0$, one can use the
geodesic distance or one can use the distances in some flat
embedding space.

Using such  a       construction, one can give the relation between the
continuum
metric $g_{\m \n}$ and the link lengths of the Regge skeleton,
\bea
 l_{1i} &=& \sqrt{ g_{11} } \; a_0 + O(a_0^2) , \nnn
 l_{2i} &=& \sqrt{ g_{22} } \; a_0+ O(a_0^2)  , \nnn
 l_{3i} &=& \sqrt{ g_{11} + 2g_{12} + g_{22} } \; a_0+ O(a_0^2) ,
          \label{LTOG}
\eea
where $a_0=1/N$ is the lattice distance on the unit torus $T_0$.
The metric tensor $g_{\m \n}$ in (\eq{LTOG}) is evaluated at the site
$x_i =(n_1 a_0, n_2 a_0)
\in \L$ and the three links $l_{1i}$, $l_{2i}$ and $l_{3i}$
are the edges on $T$ that are associated with the coordinate pairs
$\{(n_1 a_0, n_2 a_0)$, $((n_1+1) a_0 ,n_2 a_0)\}$,
$\{(n_1 a_0, n_2 a_0), (n_1 a_0 ,(n_2+1) a_0)\}$ and
$\{(n_1 a_0, n_2 a_0)$, $((n_1+1) a_0 ,(n_2+1) a_0)\}$, respectively (see
fig. 1).

Let us now for a moment take the more common point of view
and consider the  skeleton  we have constructed above
as a piecewise flat manifold, and not as  a discretization  of a continuum
manifold.
The curvature is then concentrated at the vertices  and is given by
$\sum_i \d_i \d (x-x_i)$ \cc{Regge},
where $\d_i$ is the deficit angle
at the vertex $x_i$.  The deficit angle $\d_i$ is given
by $\d_i=2 \pi-\sum_t \phi_t$, where $\phi_t$, $t=1,\ldots,6$ are the six
dihedral angles that meet at site $x_i$.
{}From the point of view that a skeleton represents an approximation of
an underlying smooth continuum manifold, such a singular curvature
distribution is unnatural. Rather we should approximate the curvature
in a region $A_i$ around the vertex $x_i$ by a constant $R_i$, such
that $R_iA_i=\int_{A_i} d^2x\sqrt{g(x)} R(x)$. This implies that
\be
R_i=2 \d_i /A_i \;.  \lb{RDA}
\ee

There is no unique prescription to assign
an area $A_i$ to a  vertex $x_i \in T_0$. The easiest way is  to
follow ref.~\cc{HaWi84} and use  a baricentric subdivision where one
divides each triangle on $T$ along the side bisections into three
equal pieces. The area $A_i$ associated with the vertex $x_i$ is then given by
\be
A_i=\third \sum_{t=1}^6 A_{it}\;, \lb{VAREA}
\ee
where $A_{i t}$, $t=1,\ldots,6$
are the six triangle areas  connected to the vertex  $i$.
For a given triangle with edge lengths
$l_\a$, $l_\b$ and $l_\g$, the triangle  area $A_{\a\b\g}$  and the
dihedral angle $\phi_\g$ which is opposite  to the edge $l_{\g}$  are
given by
\bea
 A_{\a\b\g}\;\;&=&\;\; \quart \sqrt{ -l_\a^4 - l_\b^4-l_\g^4
    +2(l_\a^2  l_\b^2 + l_\b^2 l_\g^2 + l_\a^2 l_\g^2) }
          \lb{area} \\
 \phi_{\g}\;\;&=&\;\; \arccos \left(
\frac{l_\a^2+l_\b^2-l_\g^2}{2 l_\a l_\b} \right) \;.\lb{dihedral}
\eea
With the replacement $R(x) \ra 2\d_i/A_i$
for $x\in A_i$
it is also straightforward to discretize the $R^2$ term
\cc{HaWi84} which poses a problem on a piecewise flat manifold
because  one cannot square the delta functions.
This leads to the Regge form of the action (\eq{SCONT}),
\be
S(l)= \sum_i \left\{ \l A_i + 2\k  \d_i + 4\b \d_i^2/A_i \right\} \;,
                   \label{REGGE}
\ee
with $2\sum_i \d_i=4 \pi \c$ the discrete version of the
Gauss-Bonnet theorem.
For a given torus with a sufficiently smooth metric field, this
discretized Regge action approaches the continuum action (\eq{SCONT}) in
the limit that the regularization is removed.
It can be checked, for instance using the Mathematica package, that
$A_i = \sqrt{g(x_i)}a_0^2 + O(a_0^3)$ and $R_i = R(x_i) + O(a_0)$,
when expressing the edge lengths in eqs.~(\eq{RDA}) and
(\eq{VAREA}) in terms of the metric, using relations (\ref{LTOG}),
and expanding to leading order in $a_0$.

The above discretization procedure is not invariant
under general coordinate transformations on $T$.
Let us illustrate
this for the  cosmological term. Using the relations (\eq{LTOG}) and
(\eq{area})
we can  express the sum in (\eq{VAREA})
in terms of $g_{\m\n}$,
\bea
 A_i
 \!\!&=&\!\! a_0^2 \;\sqrt{g} + \frac{a_0^3}{4 \sqrt{g}}  \left\{
                 -g_{11} \;\del_{x_2} g_{22}- g_{22}\del_{x_1}g_{11}
                     \right. \nnn
 \!\!& &\!\!         \left.
                    +  g_{12} (\del_{x_1}g_{22}+\del_{x_2}g_{11})
                    + 2g_{12}(\del_{x_1}+\del_{x_2})g_{12}
                    \right\} + O(a_0^4) \;, \lb{JUNK}
\eea
with the metric and its derivatives evaluated at the point $x_i=(n_1,n_2)$.
The relation (\eq{JUNK}) shows that the general coordinate or
diffeomorphism invariance is
broken by the $O(a_0)$ terms in the action.  This implies that
a given set of continuum manifolds
which are related by general coordinate transformations on $T$ will
lead to  Regge skeletons whose actions differ by $O(a_0)$ effects.

There are two sources of discretization effects when translating a
continuum manifold into a Regge skeleton in the above way:
Firstly, of course, discretization effects depend on the value of
$N$. Secondly, the discretization effects depend on the
choice of coordinates on $T$. By choosing an optimal coordinate system,
one can minimize the discretization effects at a given $N$.
For instance, one could choose a gauge fixing condition that minimizes the
square of the terms in the action which are
$\propto a_0^3$ (cf. eq. (\ref{JUNK})).

According to the picture sketched above, the edge lengths in a Regge skeleton
represent both physical degrees of freedom, corresponding to curvature
fluctuations, and gauge degrees of freedom, corresponding to coordinate
reparametrizations.  Since in general there does not exist
a transformation which is an invariance of the Regge action (\eq{REGGE}), the
identification of a ``gauge transformation'' on the edge lengths is
necessarily ambiguous.
For instance,  such gauge transformations on $l$  may be defined as
\bea
  l_1 \!\! & \ra & \!\! \left[l_1^2 + 2\dpp_1\x_1 \right]^{1/2},\nnn
  l_2 \!\! & \ra & \!\! \left[l_2^2 + 2\dpp_2\x_2 \right]^{1/2},\nnn
  l_3 \!\! & \ra & \!\! \left[l_3^2 + 2\dpp_1 \x_1 + 2\dpp_2 \x_2
                           +2(\dpp_1 \x_2 + \dpp_2 \x_1) \right]^{1/2} \;,
       \label{REGGAU}
\eea
where $\dpp_\m\x_\n=(\x_{\n,i+\mh}-\x_{\n,i})/a_0$. This transformation is
constructed such that after inserting relations (\eq{LTOG}) into (\eq{REGGAU}),
one recovers from the leading order terms in $\x$ the infinitesimal
continuum gauge transformations, $g_{\m\n} \ra g_{\m\n}+\del_\m
\x_\n + \del_\n \x_\m$.
The transformations (\eq{REGGAU}) are associative, i.e.
$l^{\x+{\xi}^{\pr}}=(l^{\x})^{\x^{\pr}}$. This is an important property since
it
allows  us to devide the configuration
space of Regge skeletons into equivalence classes $l^\x$ (gauge orbits),
that consist of skeletons which are related by a gauge transform of the
form (\ref{REGGAU}).
\subsection{Integration measure}
Next we turn to the problem of defining the measure for the
integration over Regge skeletons in the path integral. In the continuum
path integral, the measure (before gauge fixing) is of the form
\be
\cD g = \prod_x g(x)^{3\a} \prod_{\m\geq\n}dg_{\m\n}(x) \;.
\ee
For $\a=-1/2$
this measure is scale invariant, for $\a=-1/4$ it is gauge invariant
as follows from Fujikawa's method \cc{Fu83}.
One could now use the relation (\ref{LTOG})
to regularize this measure by expressing it in terms of the edge lengths $l$,
\be
g(x)^{3\a} \prod_{\m\geq\n}dg_{\m\n}(x)
\ra d\m(l)= 
              \half [2 A_{123}(l)]^{3\a} \prod_{\b=1}^3 dl_\b^2 \;,
\ee
where we have suppressed the site index $i$.
The triangle area
$A_{123}$ is defined in (\ref{area}).
Such a measure includes the gauge degrees of freedom.
In the continuum the gauge degrees of freedom are removed by
gauge fixing and
the full measure would acquire an additional gauge fixing and
Faddeev-Popov term. For the integration over Regge skeletons one is
not forced to fix the gauge because the gauge invariance is broken
by the the discretization procedure
and no infinite volume of the diffeomorphism group can arise.

At this point one can follow two strategies:
One can ignore the gauge
degrees of freedom (the $\x$'s of eqs. (\ref{REGGAU})) and hope that
these modes decouple automatically from the physical degrees of freedom when
approaching the continuum limit. Small $\x$ fluctuation are indeed almost
decoupled, when the Regge skeleton is sufficiently smooth as
has been demonstrated in ref. \cc{RoWi81}, but
for a generic skeleton they are strongly coupled and it is difficult to
believe that they can ever decouple from the physical degrees of freedom
in a non-perturbative evaluation of the path integral.
For  a discussion of this scenario of dynamical gauge symmetry
restoration in the context of a non-perturbative
formulation of chiral gauge theories on the lattice, see ref.~\cc{BoSm94}.

The second strategy one could follow, is to mimic the continuum
approach. This implies that the gauge degrees of freedom should be
removed from the path integral by employing a suitable gauge fixing and
Faddeev-Popov procedure. To implement this in the Regge path integral,
one could translate the continuum gauge fixing term and corresponding
Faddeev-Popov ghost term into a Regge form, e.g. using the correspondence
(\ref{LTOG}).
In the continuum the BRST symmetry prevents that gauge non-invariant
terms emerge in the low energy theory; on the lattice this symmetry is
not present and one expects that gauge non-invariant counter terms have
to be added in order to recover a BRST invariant low energy theory.
Such an approach has been advocated some time ago  for the
lattice regularization of chiral gauge theories \cc{BoMa89}, which also
have the difficulty that the local gauge symmetry is broken by the
regularization.
The technical problem with this approach is that the ghost term in the
action is very hard to include in a numerical Monte Carlo evaluation
of the path integral and a non-perturbative computation of the
counterterm coefficients may be very cumbersome.

This second option appears to have the best chance of leading
to a well-defined path integral, since it should work at least in perturbation
theory. However, faced with the technical difficulties of this strategy,
previous studies of the Regge path integral have resorted to
the first (non-gauge fixing) approach. The integration measure
was  chosen to be of a very simple form
\be
 d\m_\z(l) = \prod_{\a , i}
              \frac{dl_{\a i}^2}{l_{\a i}^2} \; l_{\a i}^\z \;,
                        \label{REGMEAS}
\ee
which is scale invariant for $\z=0$.
For a recent investigation of the $\z$ dependence of expectation values of
local observables see ref.~\cc{BeGe94}.
\section{Regge scaling relations}
Consider now the path integral for Regge skeletons with a fixed area $A$,
\be
Z(A,N_1) = \int d\m_\z(l)\;F(l)\;\exp \left[-\sum_i(\l \; A_i + \b \;
R^2_i)\right]
                 \d(\sum_i A_i-A) \;, \label{ZREGCON}
\ee
where we use the abbreviation $R_i^2=4\d^2_i/A_i$.
The total curvature of a skeleton is $4\p\c$
and we leave this term out from the action since we shall
study only skeletons with a fixed topology. The measure
$d\m_\z(l)$ is of the simple form (\ref{REGMEAS}) and
is scale invariant for $\z=0$. The factor $F(l)$ is equal to zero if
the skeleton contains one or more triangles for which
the triangle inequalities are violated, and is one otherwise.
We have in (\eq{ZREGCON}) included
the argument $N_1$, which is the number of edges (1-simplices)
in the skeleton, to
indicate that we evaluate the path integral for a fixed number of
edges with fixed connectivities.

To eliminate the delta function, we first perform a transformation
of variables,
\be
l_{\a i}\ra \sqrt{s} \; l_{\a i} \;. \label{SCALE}
\ee
Since  $F(l)$ which imposes the triangle inequalities
is unaffected by
this change of integration variables, eq.~(\eq{ZREGCON}) can be
written in the form
\be
Z(A,N_1) = \int d\m_\z(\sqrt{s}l) \; F(l) \;
          \exp \left[-\sum_i(\l \; s \; A_i + \frac{\b}{s} \; R^2_i)\right]
               \d(s \; \sum_i A_i- A) \;. \label{ZS}
\ee
Since the partition function $Z(A,N_1)$ does not depend on $s$
we may integrate both sides in (\eq{ZS})
over $s$ with a normalized weight function
$P(s)$, i.e. $\int_0^\infty ds P(s) = 1$,
\bea
 && \!\!\!\!\!\!\!\!\!\!\!\! Z(A,N_1)  = \int d\m_\z(l) F(l) \int ds
s^{N_1\z/2}P(s)
  \exp \left[-\sum_i(\l sA_i + \frac{\b}{s} R^2_i)\right]
                  \d(s\sum_i A_i-A) \nnn
&& \!\!\!\!\!\!\!\!\!\!\!\! =\exp[-\l A] \int d\m_\z(l) F(l)
          \frac{1}{\sum_i A_i} \left(\frac{A}{\sum_i A_i}\right)^{N_1\z/2}
          P\left(\frac{A}{\sum_i A_i}\right)
   \exp \left[ - \frac{\b}{A}\sum_j A_j \;\sum_i R^2_i \right] \;
\nnn
&& \!\!\!\!\!\!\!\!\!\!\!\!  \equiv
 \int d\m_\z(l) \; F(l) \; \exp [-S^{{\rm scale}}(l)]  \;,
        \label{ZSCALE}
\eea
where the last equality defines the rescaled action $S^{{\rm scale}}(l)$.

Now it appears as if $Z(A,N_1)$ depends on the arbitrary function $P$, but this
is of course not the case. This follows from  an infinite
set of Ward identities which can be derived by using again the scale
transformation
(\ref{SCALE}), and observing that $d^n Z(A,N_1)/ds^n=0$, $n=1,2,3,\ldots$.
For $n=1$ this gives the identity
\be
 \left\langle \frac{A}{\sum_iA_i} P^{\prime}\left(\frac{A}{\sum_iA_i}\right)
                       \left/ \right.
          P\left(\frac{A}{\sum_iA_i}\right) \right\rangle  = -1\;,
            \label{PID}
\ee
where the expectation value is with respect to the rescaled action,
\be
\av{\bullet} = Z(A,N_1)^{-1}\int d\m_\z(l)  \;F(l) \; \bullet  \;
  \exp [-S^{{\rm scale}}(l)]  \;.
\ee

We want to investigate if the Regge partition function $Z(A,N_1)$ reproduces
the scaling behavior (\ref{ZCONT}) in the continuum limit,
i.e. for $N_1 \ra \infty$. Therefore it is useful to consider the derivative
\be
\frac{d \log Z(A,N_1)}{dA}= -\l + \frac{1}{A}\left(
\frac{\b}{A}\av{\sum_jA_j\sum_iR^2_i } + \frac{N_1\z}{2} -1 \right) \;,
                \lb{DZDA}
\ee
with $\l \b$ and $N_1$ kept fixed. Here we have used the identity
(\ref{PID}) to eliminate a $P^{\prime}/P$ term.

Next we note that the expectation value $\av{\sum_jA_j \; \sum_iR^2_i }$
is a function of the ratio $\b/A$ and of $N_1$. This follows from
inspecting the path integral (\eq{ZSCALE}).
If for $N_1\ra\infty$ this expectation value behaves such that
\be
 \frac{\b}{A} \; \av{\sum_jA_j \; \sum_iR^2_i } + \frac{N_1\z}{2}
              = -\frac{A}{\b} \; \d\l +\d\g  +O(\frac{\b}{A}) \;, \label{HOPE}
\ee
then the Regge partition function would scale as
\be
 Z(A,\infty)\propto \exp \left[- \left(\l+\frac{\d\l}{\b} \right) A \right]
A^{-1+\d\g}
           \lb{HOPE2}
\ee
and we may identify $\l+\d\l/\b$ with the renormalized cosmological
constant and $\d \g+2$ with the string susceptibility $\g_{{\rm str}}$.

Since the term  $(\b/A)\sum_j A_j \;\sum_iR^2_i $ is the total action
in the path integral (\eq{ZSCALE}), its  expectation value
in general will grow proportionally to $N_1$. Hence the limit
$N_1\ra\infty$ only exists, if the exponent $\z$ in the measure is chosen such
that $N_1\z/2$ cancels the term $\propto N_1$ in the expectation value of
the action.

Instead of the measure (\eq{REGMEAS}) we could also
have used a different measure where the exponent $\z$ is given by
$\z=\z_0 + \z_1/N_1$. Now the term $N_1 \z_0/2$
has to cancel the divergence, but the finite part is shifted by the
arbitrary constant $\z_1$ which gives a contribution to the string
susceptibility. Hence $\d\g$ appears to depend very sensitively on the choice
of the measure. The only way that $\d\g$ could be independent of $\z_1$
is if it is compensated by the expectation value of the action:
An expansion to lowest order in $\z_1/N_1$
shows that $\g_{{\rm str}}$ is shifted by
$\z_1(\half+c)$ with $c$ the connected part of
$\av{(\sum_{\a , i}\log l_{\a i})(\sum_j A_j\sum_i\b R_i^2/A)}$, evaluated
for $\z_1=0$.
It is very unlikely that $c$ is exactly equal to $-1/2$, such that
$\d\g$ is independent of $\z_1$.

This sensitive dependence of $\d\g$ on the measure is seen particularly
clearly for $\b=0$:
Then eq. (\ref{DZDA}) can be integrated easily, giving
\be
  Z(A,N_1) \propto A^{-1+N_1 \z_0/2+\z_1/2} \; \exp[-\l A] \;.   \lb{ZBZERO}
\ee
For the choice  $\z_0=0$ the limit $N_1\ra \infty$ can be taken
and the partition function is of the same
form as in the continuum, but with an arbitrary string susceptibility
$\g_{{\rm str}}=2+\z_1/2$.

This discussion shows that the real difficulty is the lack of a criterion to
fix such ambiguities. This is different in the continuum where the power
$\z$ is fixed in Fujikawa's method \cc{Fu83}, by the
requirement of gauge invariance.
Such a guideline is lacking in the Regge model.

These considerations cast serious doubts on the claim made in ref. \cc{GrHa91}
that the Regge model with scale invariant measure reproduces the
continuum values of the string susceptibility. To settle this, we
have also performed a numerical simulation of this model
using the scale invariant measure, i.e. for $\z=0$, and
three different topologies with genus $h =0,1$ and $2$. We
shall see that the numerical estimates for $\g_{{\rm str}}$ differ
substantially from the continuum result for $h\ne 1$.
\section{Details of the numerical simulation}
\subsection{Histogramming}
The area dependence of the Regge partition function can be inferred from
eq. (\ref{DZDA}) by computing
$\langle (\b/A)\sum_jA_j\sum_iR^2_i \rangle =-\bt d\log Z(A,N_1)/d\b$
in the rescaled model (\eq{ZSCALE}).
This is identical to evaluating the expectation value of $\b R^2$ in the
original model with the delta function
(\ref{ZREGCON}).  Hence there are two ways to proceed:
One can simulate either the original model, including the delta function that
constraints the area to be fixed,  or alternatively
one can simulate the rescaled model (\eq{ZSCALE}).

The first option poses the problem of including the constraint that the
total area must be fixed. Such a global constraint is hard to implement
efficiently in a Monte Carlo simulation (see however ref.~\cc{Be85}).
Alternatively it is possible to
include the constraint a posteriori. To this extend we  first approximate the
delta function by a narrow block function which is normalized to one,
\be
 \D_h(A)    =\left\{ \begin{array}{ll}

\frac{1}{2h} \;,\;\;\;\; &  \mbox{if} \;\;A       \in [-h,h] \\
      0      \;,\;\;\;\; &  \mbox{if} \;\;A  \not \in [-h,h] \;.
                  \end{array} \right.
\ee
Then the expectation value of the observable $O(l)$ is
given by,
\bea
  \av{O}_A \!\!& = &\!\! Z(A,N_1)^{-1}\int \; d\m_\z(l) \; F(l) \; O(l)\;
     \D_h (\sum_iA_i-A)
\; \exp [-S(l)] \nnn
 \!\!& = &\!\! \av{O(l)\; \D_h (\sum_iA_i-A)}/\av{\D_h (\sum_iA_i-A)}\;,
\label{CONEX}
\eea
where the action $S(l)$ is given in eq.~(\eq{REGGE}) with $\k=0$
and $\langle \bullet \rangle$
in the lower line is evaluated in the
model without the area constraint. In practise this amounts to generating
an equilibrium ensemble of skeletons with the unconstraint action and
then computing the average of $\b R^2$ on the subensemble for which the
total area is within the interval $[A-h,A+h]$ for a selected value of $A$.
We shall refer to this method as ``the histogramming technique''.

Most of our numerical results to be presented below, have been
obtained with this method. For the simulation of the unconstraint action
we use a Metropolis algorithm. The $R^2$ term in the Regge action
extends over several triangles
and in order to make the code vectorizable we had to use
a four color subdivision of the lattice.
In the model without the area constraint
we can derive the Ward identity
\bea
 && \!\!\!\!\!\!\!\!\! -\frac{N_1\z}{2} + \frac{1}{Z}
\int d \m_\z \; F(l) \; \left( \l \; \sum_i A_i-\b \; \sum_i R^2_i \right)
\exp[-S(l)]= 0 \;, \lb{WARD} \\
&& \!\!\!\!\!\!\!\!\! Z=\int d \m_\z \;   F(l) \; \exp[-S(l)] \;,
\lb{ZWARD}
\eea
which follows from the invariance under rescaling of the links.
This shows that the unconstraint path integral is divergent for
$\z=0$ and $\b=0$ because  the expectation value of
$\sum_i A_i>0$ can only vanish if (\eq{ZWARD}) diverges.
This is perhaps seen even more clearly by writing the unconstraint
partition function in the form $Z=\int_0^{\infty} dA \; Z(A, N_1)$
with $Z(A,N_1)$ given in eq.~(\eq{ZREGCON}).
The  integration of relation (\ref{DZDA}) for $\b=0$ yields
$Z(A,N_1) \propto A^{\z/2 -1} \; \exp[-\l A]$, which shows that
the unconstraint partition
function $Z$  has a non-integrable singularity at $A=0$
if  $\z\leq 0$.

Also the rescaled action (\ref{ZSCALE}) could be simulated using a
Metropolis algorithm.
However, the non-local action $\b \sum_j A_j\sum_iR^2_i$
prevents vectorization of the code. This suggests to
use a Hybrid Monte Carlo algorithm, which can easily be vectorized
and is usually found to be more efficient than a Metropolis algorithm.
\subsection{Hybrid Monte Carlo algorithm}
To use the Hybrid Monte Carlo algorithm  we first
introduce new degrees of freedom
$p_{\a i}$ which are viewed as the canonical momenta
of  the edge lengths $l_{\a i}$ with respect to the measure
$dl_{\a i} \; dp_{\a i}$.
The path integral (\eq{ZSCALE}) may then be written in the form
\bea
Z\!\!&= &\!\! \int \prod_{\a , i}
      (dl_{\a i} \; d p_{\a i}) \; F(l) \; \exp[-H(l,p)] \;, \lb{PHMC} \\
   H(l,p)   \!\!&=&\!\! \half\sum_{\a , i}p_{\a i}^2 + \hat{S}(l) \;, \lb{HHMC}
\\
 \hat{S} (l) \!\!&=&\!\!  \frac{\b}{A} \; \sum_j A_j \; \sum_i R^2_i -
                    \ln P \left( \frac{A}{\sum_i A_i} \right) \nonumber \\
  \!\!& &\!\!       + (\half  \; N_1  \; \z   +1) \;
                    \ln \left( \sum_i A_i \right) -
                    (\z-1) \; \sum_{\a , i} \ln l_{\a i} \;, \lb{SHMC}
\eea
where we have introduced the hamiltonian $H(l,p)$ and dropped the constant $\l
A
- \half N_1\z \ln A$ from $\hat{S}(l)$.
The various factors in (\eq{ZSCALE}) have been included as
logarithms in the action, except for $F(l)$  which
is a discontinuous function of the edge length $l$.

We now want to find
trajectories for $l(t)$ and $p(t)$ such that the hamiltonian $H(l,p)$ is
conserved
in the (computer) time $t$. After choosing the canonical relation
\be
\frac{d l}{dt} = \frac{\del H(l,p)}{\del p} =p \;, \lb{LUP}
\ee
we can compute the
time derivative of $p$ from the requirement that $d H/dt=0$,
which leads to
\be
\frac{d p_{\a i}}{ dt} = - \frac{\del \hat{S}(l)}{\del l_{\a i}} \;. \lb{PUP}
\ee

So far we have not done anything to include the constraint $F(l)$
on the edge lengths. The region in configurations space where the
triangle inequalities
are violated can be avoided by setting the action in this
subspace equal to infinity.
The trajectories then should  bounce back from this potential barrier,
with the condition that the momentum is conserved.
One way to achieve this is to flip the sign of the momentum component
perpendicular to
the plane $p_\a-p_\b-p_\g=0$, when the
triangle with edge lengths $l_\a$, $l_\b$ and $l_\g$
just starts to fail the constraint $l_\a-l_\b-l_\g>0$.
A less sophisticated procedure is to flip the signs of all
momenta, $p_\a \ra -p_\a$, $p_\b \ra -p_\b$, $p_\g \ra -p_\g$.

A new configuration $\{l_{\a i} \}$ is computed by integrating the hamiltonian
equations of motion (\eq{LUP}) and (\eq{PUP}) from $t$ to $t+\D t$.
These equations are integrated numerically by means of
the standard leap frog method \cc{HMC}, which preserves the time
reversibility of the evolution.
To compensate for discretization errors, $H(t+\D t )=H(t)+O(\d t^2)$, one
accepts the new configuration $\{l(t+\D t),p(t+\D t)\}$ with probability
$\mbox{min}\{1,\exp[ -( H(t+\D t)-H(t) ) ]\}$.
We use a discretization time step $\d t=\D t/n$, with $n\approx 10$
and $\d t$ chosen such that the acceptance rate is between 60 and 80\%.

A practical way to implement the constraint $F(l)$ is the following:
We check after
each time step $\d t$ if one or more triangle inequalities
are violated.  If this is the case, we assume that the momenta are
constant over the interval $\d t$ and compute the earliest time
when such a violation has occurred. We use one
of the above mentioned prescriptions to flip the momenta which
correspond to the links
that are responsible for the triangle inequality violation.
Afterwards we try to complete the time step with the new set of momenta and
repeat the above procedure of flipping momenta until no triangle
inequalities are violated at $t+\d t$.
This procedure, which is easy to implement in two dimensions,
ensures that the time evolution is still
reversible in $t$ when the discretized trajectory bounces off
the regions in
configuration space where triangle inequalities are violated. We find that
the number of triangle      inequality violations is
typically small and decreases at larger $\b$.
\begin{figure} [tttttbbb]
\vspace{-5.0cm}
 \centerline{ \epsfysize=7.5cm \epsfbox{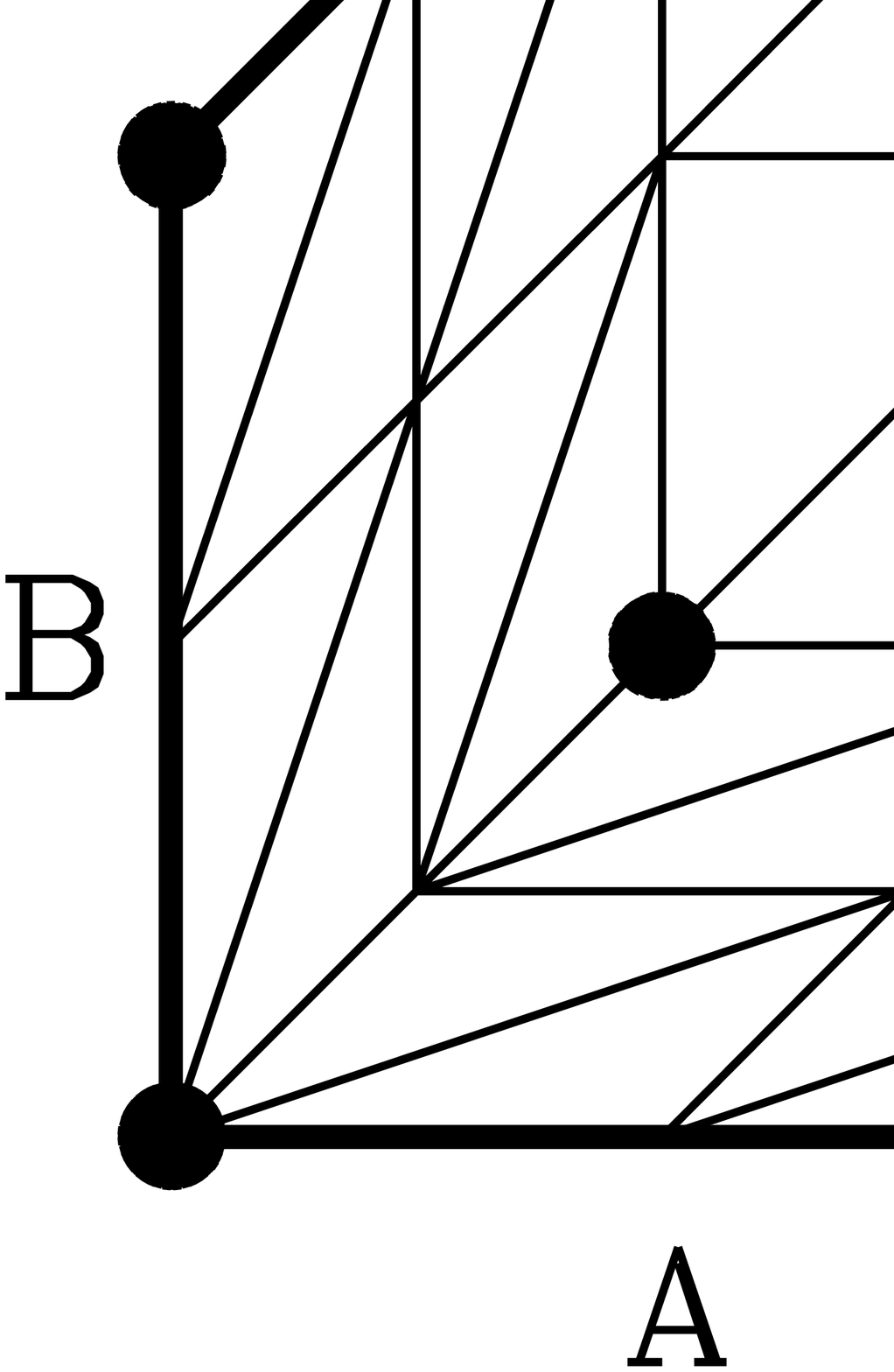} }
\vspace{-3.0cm}
 \centerline{ \epsfysize=7.5cm \epsfbox{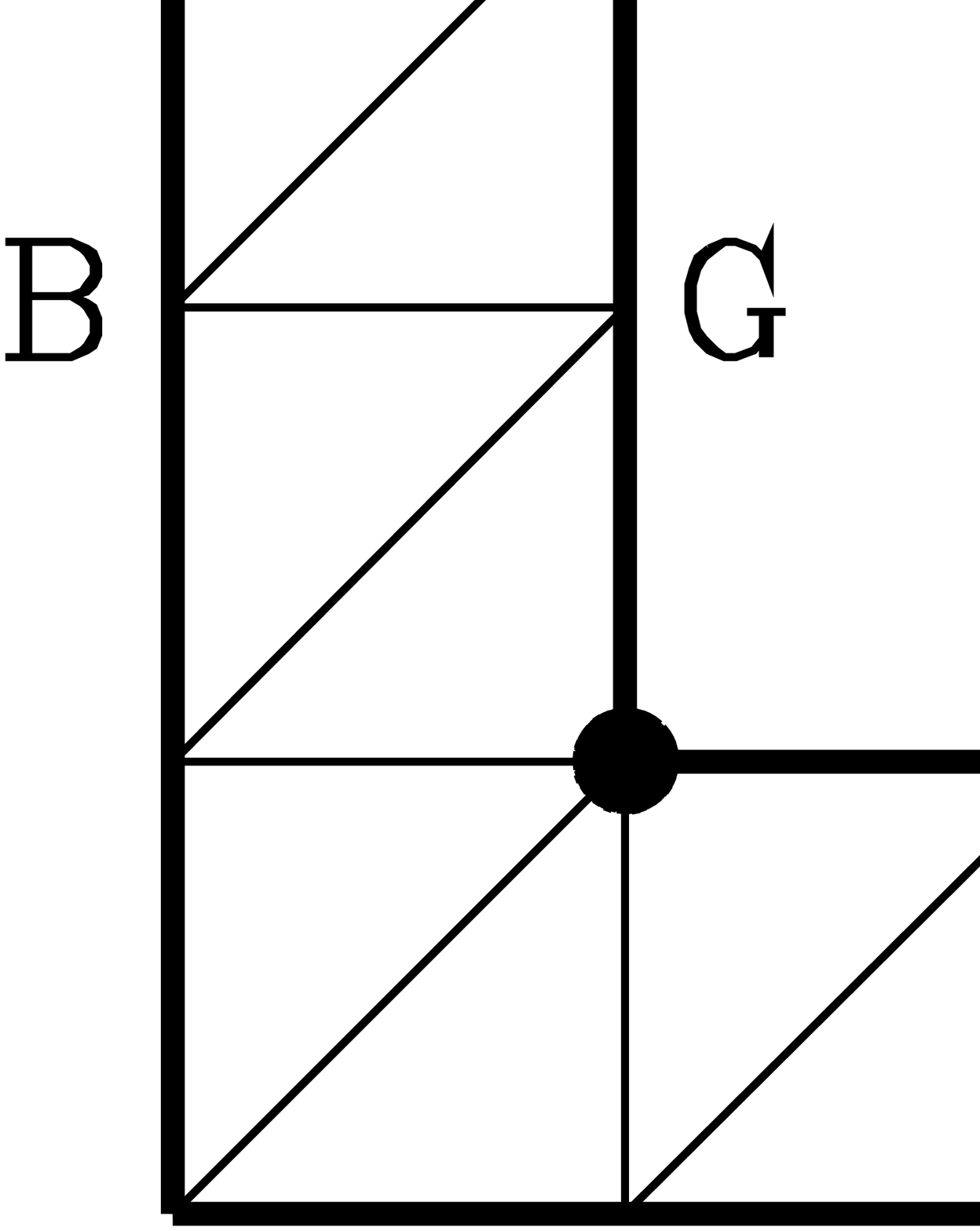} }
\vspace{ 0.2cm}
\caption{{\em Triangulation  of
              the sphere with $N=3$ (fig. a) and the bi-torus with $N=4$
              (fig.  b). The boundaries with the same labels have to be
              identified. The full circles mark the vertices whose coordination
              number differs from six.}}
\lb{fig1b}
\end{figure}

{}From eqs.~(\eq{area}) and (\eq{SHMC}) it follows that $\del \hat{S} (l)/
\del l_\a$ contains terms  of the form
$\del A_{\a\b\g}/\del l_\a = l_\a(-l_\a^2 + l_\b^2 + l_\g^2)/8A_{\a\b\g}$,
which diverge when $l_\b \ra l_\a + l_\g$ or $l_\g\ra l_\a + l_\b$.
This implies that discretization errors may blow up near the excluded
regions  in configuration space, which implies that the
time step $\d t$ has to be
taken very small in order to keep the acceptance rate sufficiently large.
The reduction of $\d t$ makes the algorithm relatively slow.
We could improve on this by generalizing the triangle
inequalities, $F(l) \ra F_{\e}(l)$, such that the minimum area
that is allowed, is increased from
zero to some small non-zero value, i.e. by using the inequalities
$-l_\a+l_\b+l_\g>\e >    0$ instead of $-l_\a+l_\b+l_\g>0$.
With this modification the Hybrid Monte Carlo algorithm becomes competitive
in speed with the Metropolis algorithm and is more
efficient for larger values of $\b$.

\subsection{Topologies with genus zero, one and two}
Let us close this technical section with a description of
the various topologies we have used in our simulations.
The simplest skeleton has the topology of a
torus, cf. fig.~\ref{fig1}. We use the same
simple triangulation as in ref.~\cc{HaWi84,GrHa91}, which is obtained
by taking a regular lattice with periodic boundary conditions
in spatial and time directions and
adding diagonal links, such that each site $x=(n_1 a,n_2 a)$ is
connected by six links, as discussed in sect.~2.2.
On a torus with $N_0=N^2$ vertices ($0$-simplices), the number or triangles
($2$-simplices) is
$N_2=2N_0$ and the number of links
($1$-simplices) is $N_1=3 N_0$. The Euler
characteristic is $\c =N_2-N_1+N_0=0$.

Our second topology is that of a sphere, which has the Euler characteristic
$\c=2$. Here we do not follow the construction of ref.~\cc{GrHa91},
in which a cylinder is made into a sphere by pinching each of the two
boundaries
to a point. These two vertices have an exceedingly large coordination number
of order $\sqrt{N_0}$ which has been claimed in ref.~\cc{GrHa91}
to affect the scaling of
the partition function. To avoid large coordination numbers,
we start from a three dimensional cube and use the two dimensional
lattice consisting of its boundary. Each of its six faces is triangulated
in the same regular way as the torus. In fig.~\ref{fig1b}a we have cut the cube
open
in order to show    the triangulation of its front and back side.
All vertices have coordination number six except for
six of the eight corner points whose coordination number is four.
If the number of sites on the square faces of the cube is $N^2$,
the total number of vertices on the sphere is $N_0=6(N  -1)^2+2$,
the number of triangles is
$N_2=2N_0-4$ and the number of links is $N_1=3N_0-6$.

To illustrate how typical surfaces look like, we have embedded the
two dimensional surfaces of the sphere
in a three dimensional space. For typical equilibrium configuration  at
a small and large value of the $R^2$ coupling $\b/A$
we have determined the  embedding space
coordinates $X_i^1$, $X_i^2$ and $X_i^3$ associated with vertex $i$
using the relations
$l_{\a i}^2 = \sum_{\b=1}^3 (X_i^\b-X_j^\b)^2$, with $i$ and $j$
denoting the two neighboring vertices connected by the edge $l_{\a i}$.
Fig.~\ref{FIGG} exhibits two examples for such
surfaces, one obtained at $\b/A=0.5$ (left graph) and the  other at
$\b/A=0.0005$ (right graph). It is clearly seen
that the surface  at $\b/A=0.5$ is smoother and
appears to be much closer to  a sphere, whereas the surface at $\b/A=0.0005$ is
more irregular and crumbled.
Notice that the underlying features of the cube have completely disappeared
and parts of the surface which contain
a  vertex   with coordination
number four do   not to carry an exceptional  amount  of curvature.
\begin{figure} [tttttbbb]
\vspace{-1.7cm}
\centerline{
\vspace{-3.0cm}
\epsfysize=11.5cm \epsfbox{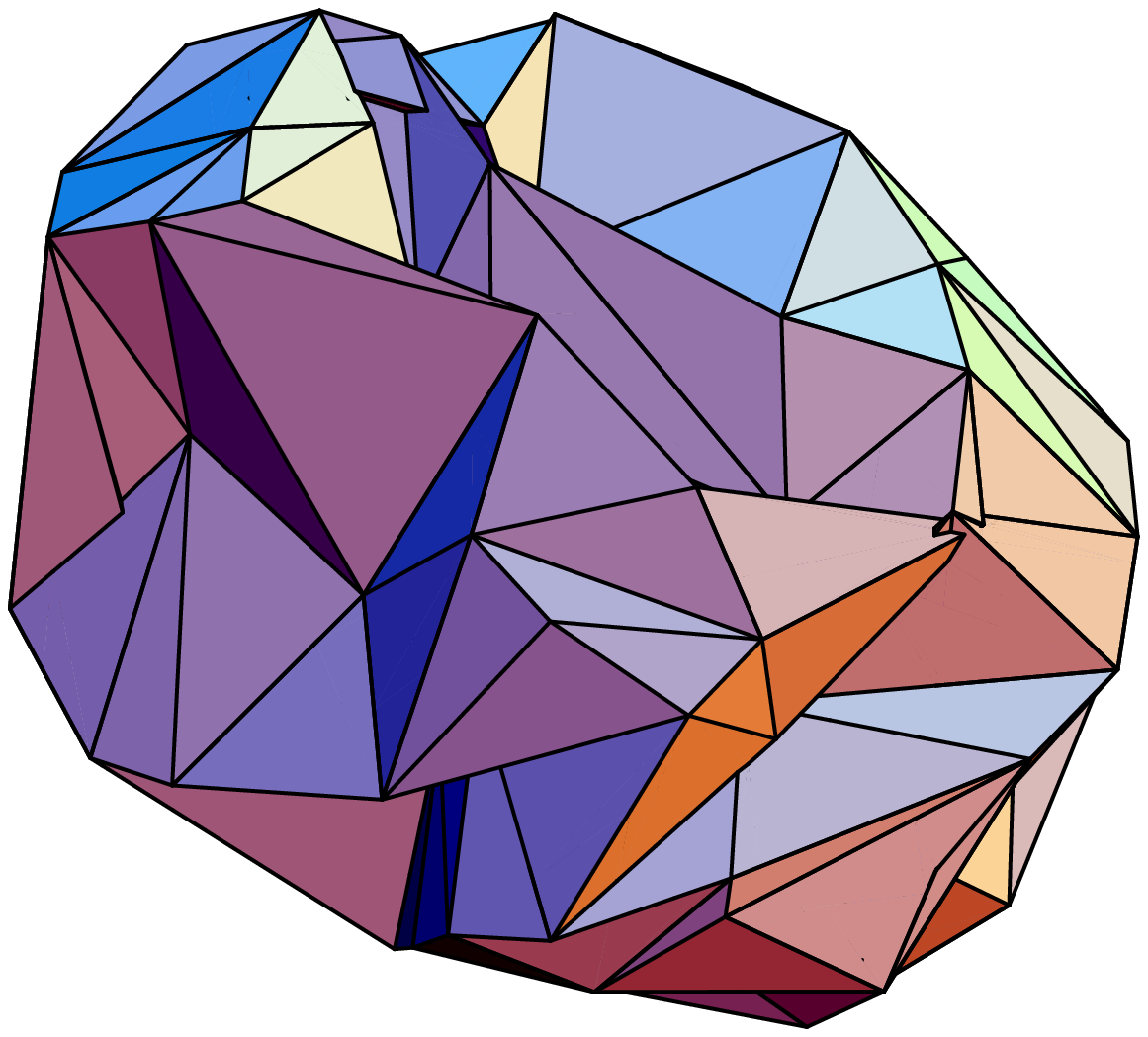}
}
\vspace{-9.0cm}
\centerline{
  \epsfysize=11.5cm \epsfbox{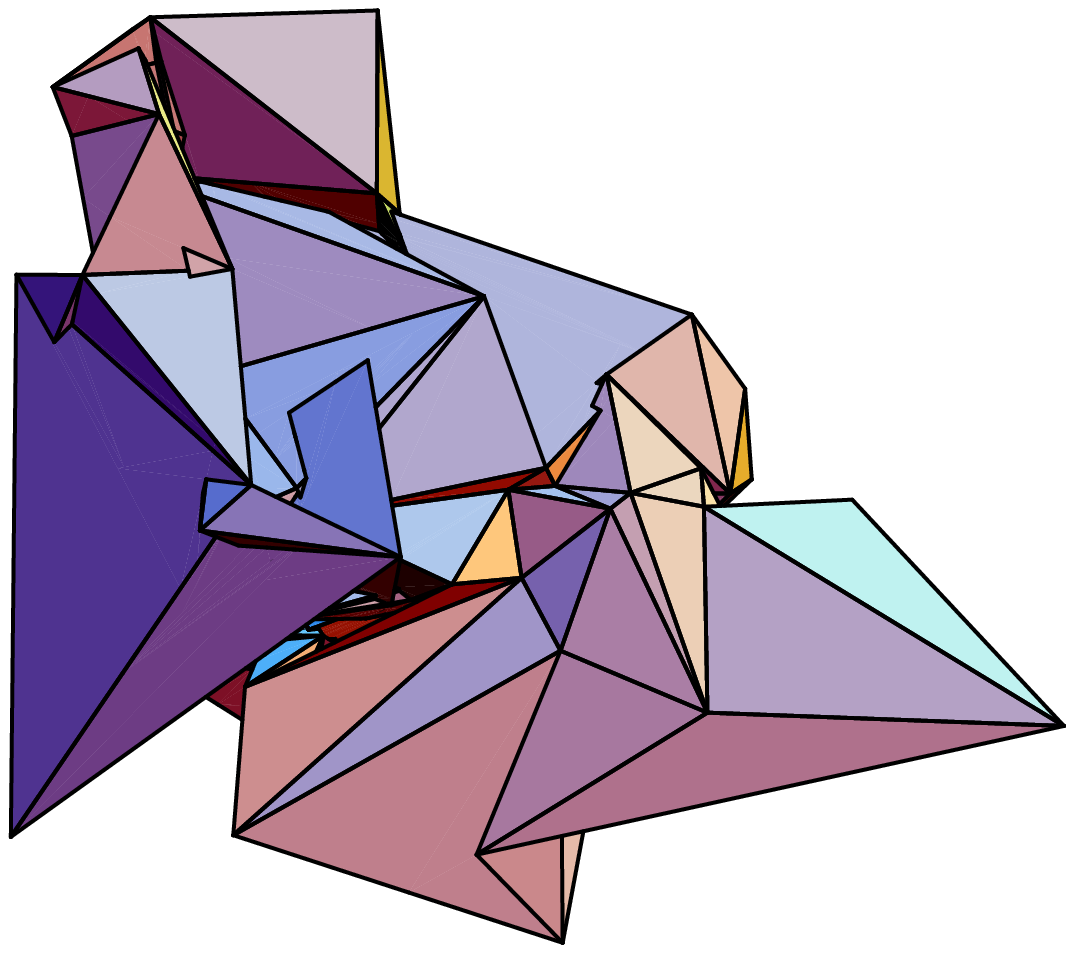}
}
\vspace{ -2.3cm}
\caption{{\em  Equilibrium surfaces with the topology of a
 sphere, embedded in three dimensional flat space.
             The $R^2$ couplings are  $\b/A=0.5$ (left graph)
  and $\b/A=0.0005$ (right graph). The number of triangles  is $N_2=192$.
 In the left graph one vertex with coordination number four is visible.}}
\lb{FIGG}
\end{figure}

Finally we have also constructed skeletons with $\c=-2$, which have two
handles. These skeletons are obtained by gluing together two tori of
size $N^2$, as illustrated in fig.~\ref{fig1b}b. In
both tori we cut out a square window of size $N/2$ and glue them together
by identifying the links that are on the boundaries of the two windows.
The four corner points in the window frame have a coordination number
larger than six: two have coordination number eight and two have ten.
All other vertices have coordination number six.
The total number of vertices on such a bi-torus is $N_0=3 N^2/2-2$
($N$ must be even), the number or triangles is
$N_2=2N_0+4$ and the number of links is $N_1=3N_0+6$.
\section{Numerical results}
In order to compute the string susceptibility and renormalized cosmological
constant we have to compute the $A/\b$ dependence of the average action
in the constraint area model for $N_2 \ra \infty$.
This term should behave as
$(\b /A) \langle \sum_j A_j \; \sum_i R_i^2 \rangle = (A/\b) \;\d\l + \d \g
+O(\b/A)$, which leads to
the string susceptibility $\g_{{\rm str}} =\d\g +2$,
cf. eqs. (\ref{HOPE}) and (\ref{HOPE2}) above.

\begin{figure} [ttb]
\vspace{-0.5cm}
 \centerline{ \epsfysize=9.5cm \epsfbox{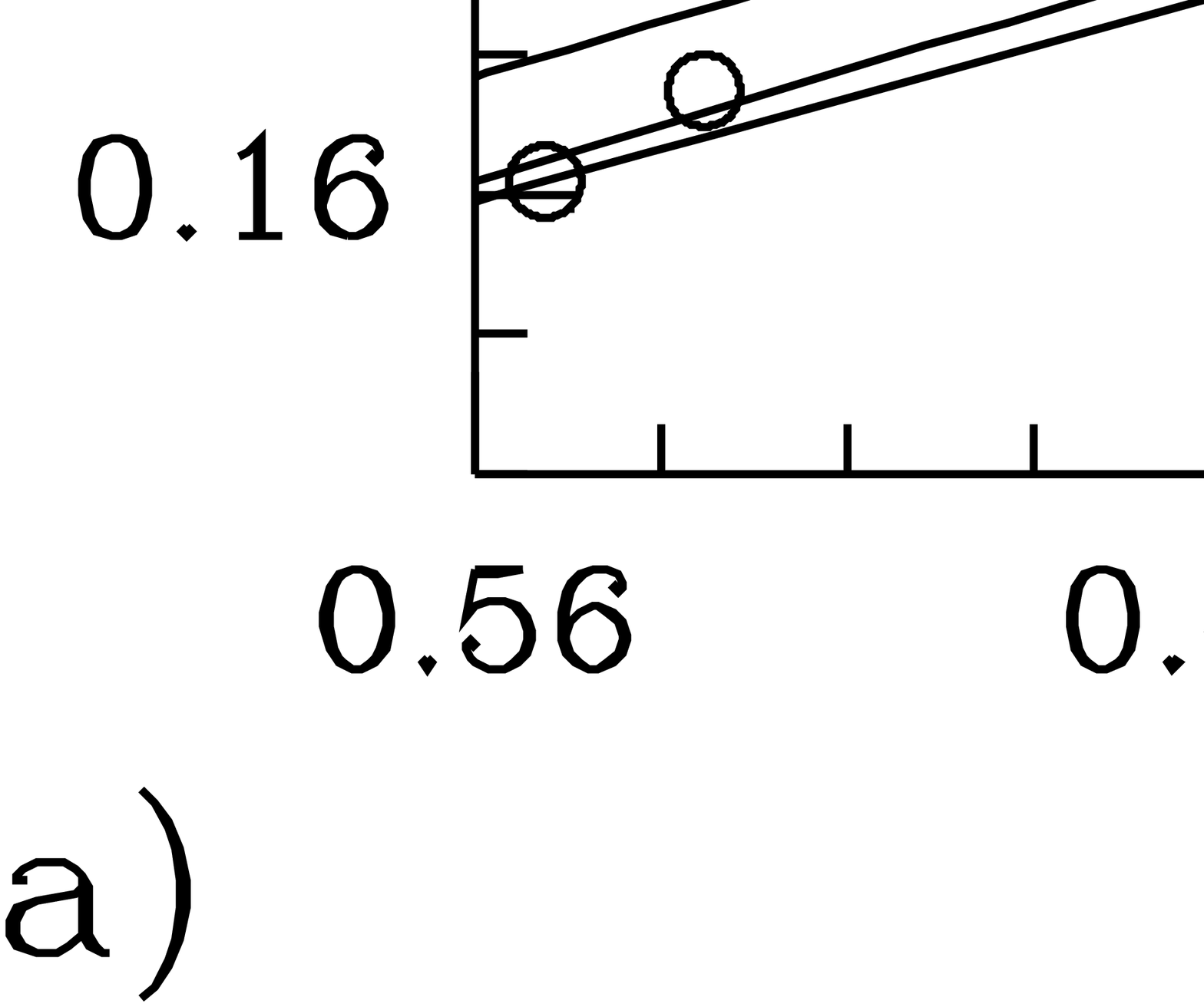} }
\vspace{0.7cm}
 \centerline{ \epsfysize=9.5cm \epsfbox{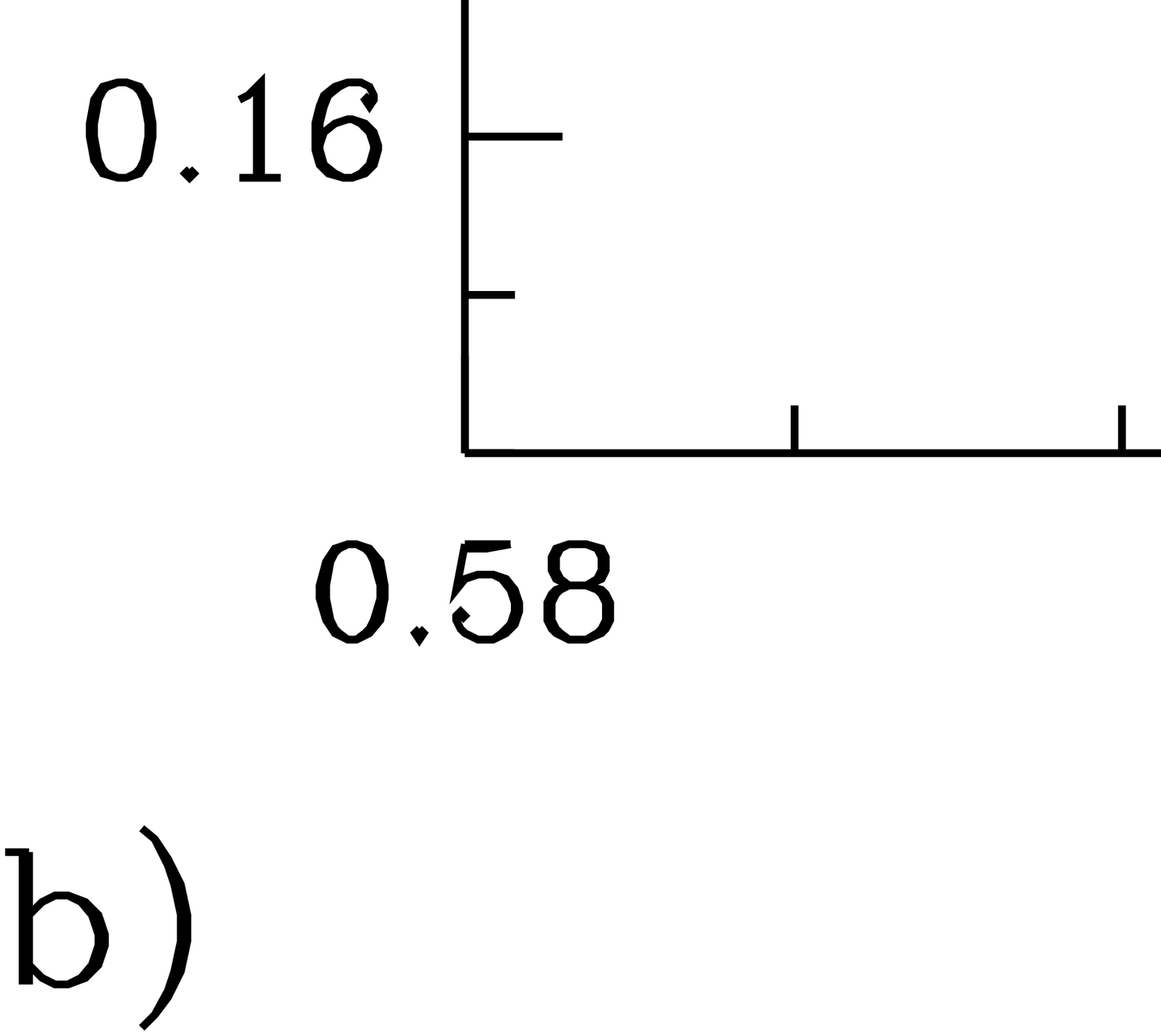} }
\end{figure}
\begin{figure} [ttt]
\vspace{-0.5cm}
 \centerline{ \epsfysize=9.5cm \epsfbox{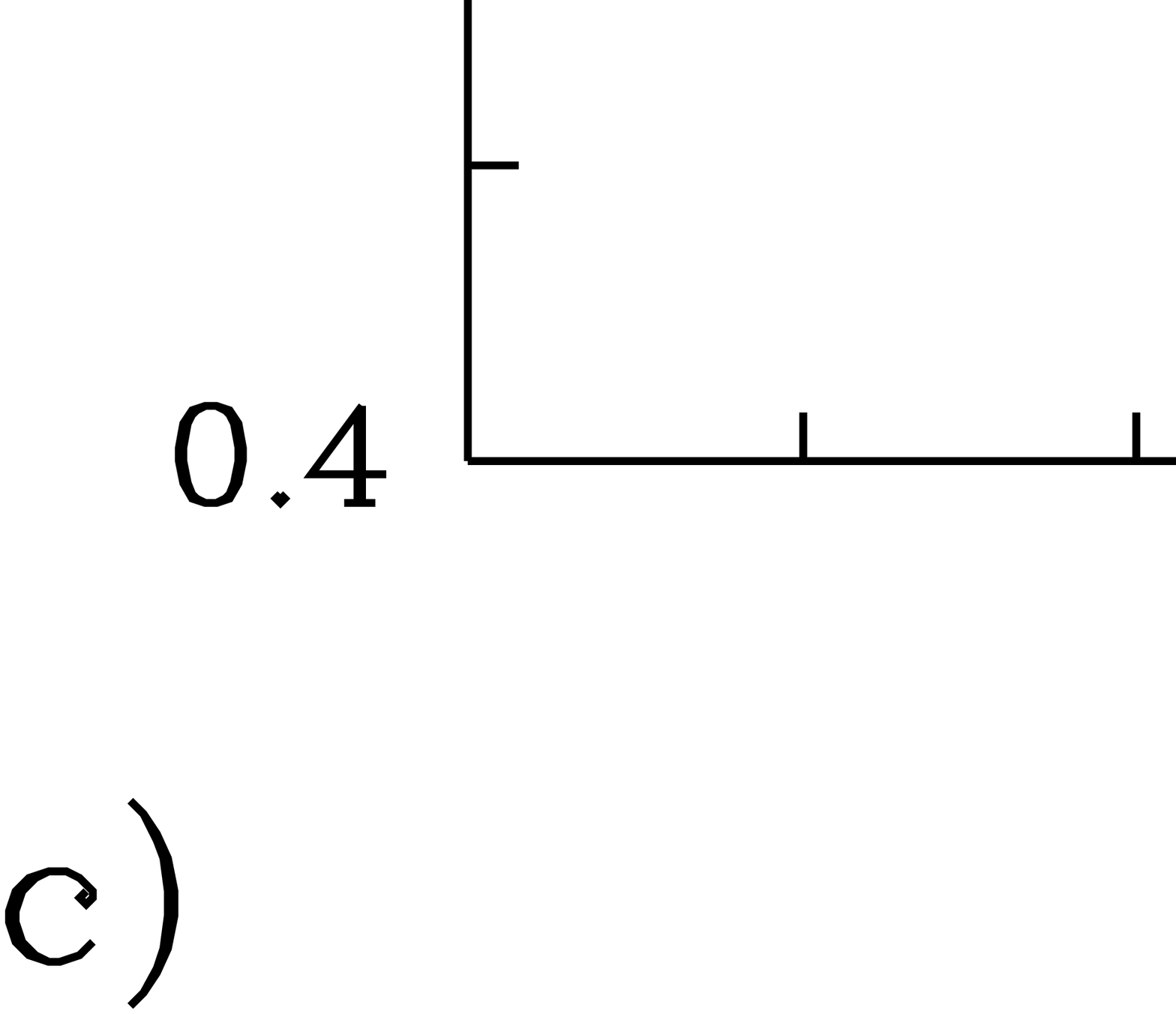} }
\vspace{-0.5cm}
\caption{{\em Histograms for the bi-torus (fig.~a), the
              sphere (fig.~b) and the torus (fig.~c).
              We have plotted
              $S_{R^2}(\b/a) = \b \av{\sum_i R_i^2/N_2}_A$
              as a function of $\b/a$ for several lattices
              (bi-torus: from top   to bottom $N_2=108, \ldots ,
              1728$;
              sphere: from top   to bottom $N_2=192, \ldots ,
              4332$;
              torus: from bottom to top   $N_2=18, \ldots ,
              4608$). The circles represent
              the data points. We have dropped the error bars when they
              are smaller than the symbol size.
              }}
\lb{fig2}
\end{figure}

Most of our data have been generated with the unconstraint action.
We used  the histogramming method to compute  the average action
$\b\langle  \sum_i R_i^2 \rangle_A$
with the constraint expectation value
$\av{\bullet}_A$ defined in eq.~(\eq{CONEX}).
For a fixed value of $\b$ the distribution of triangle areas in an
ensemble is peaked around an average triangle area
$a\equiv \av{\sum_i A_i}/N_2$, with $N_2$ the number of triangles in the
skeleton. For increasing $N_2$ the peak of the distribution
becomes more narrow, but remains roughly at the same value of $a$.
In our numerical simulations we have
kept $\b$ fixed when increasing $N_2$.
Eq.~(\eq{WARD}) shows that it then would require
tuning $\b \propto N_2$ to keep the bin with a fixed value
of $\sum_iA_i/\b$ sufficiently populated when increasing the skeleton
size.

It requires several steps to estimate a value for the shift of the
string susceptibility $\d \g$: First we generate a large
ensemble of $10^5-3 \times 10^5$ configurations for each lattice size
and store for each configuration the values of $\sum_i A_i$ and $\sum_i R^2_i$.
Then we use the histogramming technique to compute the expectation value
$S_{R^2}(\b/a) \equiv \b \av{\sum_i R_i^2/N_2}_A$
for several values of the ratio $\b/a$. In fig.~\ref{fig2} we
have plotted $S_{R^2}(\b/a)$ as a function of $\b/a$
for skeletons with topologies of a bi-torus (fig. \ref{fig2}a),
a sphere (fig. \ref{fig2}b) and
a torus (fig. \ref{fig2}c). The circles indicate the centers of the bins. On
the larger
lattices the width of the distribution is substantially larger  than the
$\b/a$ interval shown in fig.~\ref{fig2} and most of the data points are
outside the frames.
Each  figure contains the results for a number of different lattice sizes.
These distributions overlap over the full range of values of
$N_2$ which makes it easy to read of the $N_2$ dependence.
As noted above, this would not be the case if we would plot $S_{R^2}$ as a
function of $\b/\sum_i A_i$. The solid lines in figs.~\ref{fig2}a, b,
and c have been obtained  by a polynomial fit which has been used
to interpolate these results to obtain $S_{R^2}(\b/a)$ over a
range of $\b/a$ values. The fits on the smaller lattices include also those
data which are
outside the frames.

In figs.~\ref{fig3}a, b and c we have displayed  the interpolated values for
$S_{R^2}(\b/a)$ as a function of $1/N_2$ for the three different topologies
and
several fixed values of $\b/a$. It is seen that for the larger values of
$N_2$ shown in the figure, the data  fall nicely
on straight lines. For smaller $N_2$ we find deviations from the
straight line behavior.
This shows that for sufficiently large $N_2$ $S_{R^2}(\b/a)$ behaves as
\be
 S_{R^2}(\b/a) = c_0 (\b/a) + c_1(\b/a)/N_2 + \cdots \;. \lb{R2ANS}
\ee
The coefficients $c_0$ and $c_1$ are
found using a $\chi^2$ fit (straight lines in figs. \ref{fig3}a, b and c).
The value of $\chi^2$ was in all cases smaller than
two.

Next we have to convert the results for $c_0$ and $c_1$ into an estimate
for the string susceptibility and the renormalized cosmological constant.
In sect. 3 we have shown that,
\be
\av{\b\sum_iR^2}_A = N_2 \; S_{R^2}(N_2\b/A) =
N_2 \; c_0(N_2\b/A) + c_1(N_2\b/A) + \cdots
\ee
should behave as,  cf. eq. (\eq{HOPE}),
\be
\av{\b\sum_iR^2}_A
        = -\frac{N_1 \;\z}{2} - \frac{A}{\b} \; \d\l + \d\g+ O(\b/A)\;.
\ee
This is the case if the coefficients $c_0$ and $c_1$ can be expanded as
$c_{0,1}(x)=c_{0,1}^{(0)} + c_{0,1}^{(1)}/x + O(1/x^2)$.
Combining all results then leads to the relations
\be
       \d\l=-c_0^{(1)}\;,\;\; \;\;\; \;\;\; \d\g=c_1^{(0)}\;.
        \lb{LINK}
\ee
In order to cancel the term $\propto N_1$ in the right hand side of
eq. (\ref{HOPE}), we should choose the measure such that
\be
\z +4\; c_0^{(0)}/3=0 \;.  \lb{CAN}
\ee
Here we have used the relation $N_1/N_2 =3/2$ which is valid for any two
dimensional skeleton.

\begin{figure} [ttb]
\vspace{-0.5cm}
 \centerline{ \epsfysize=9.50cm \epsfbox{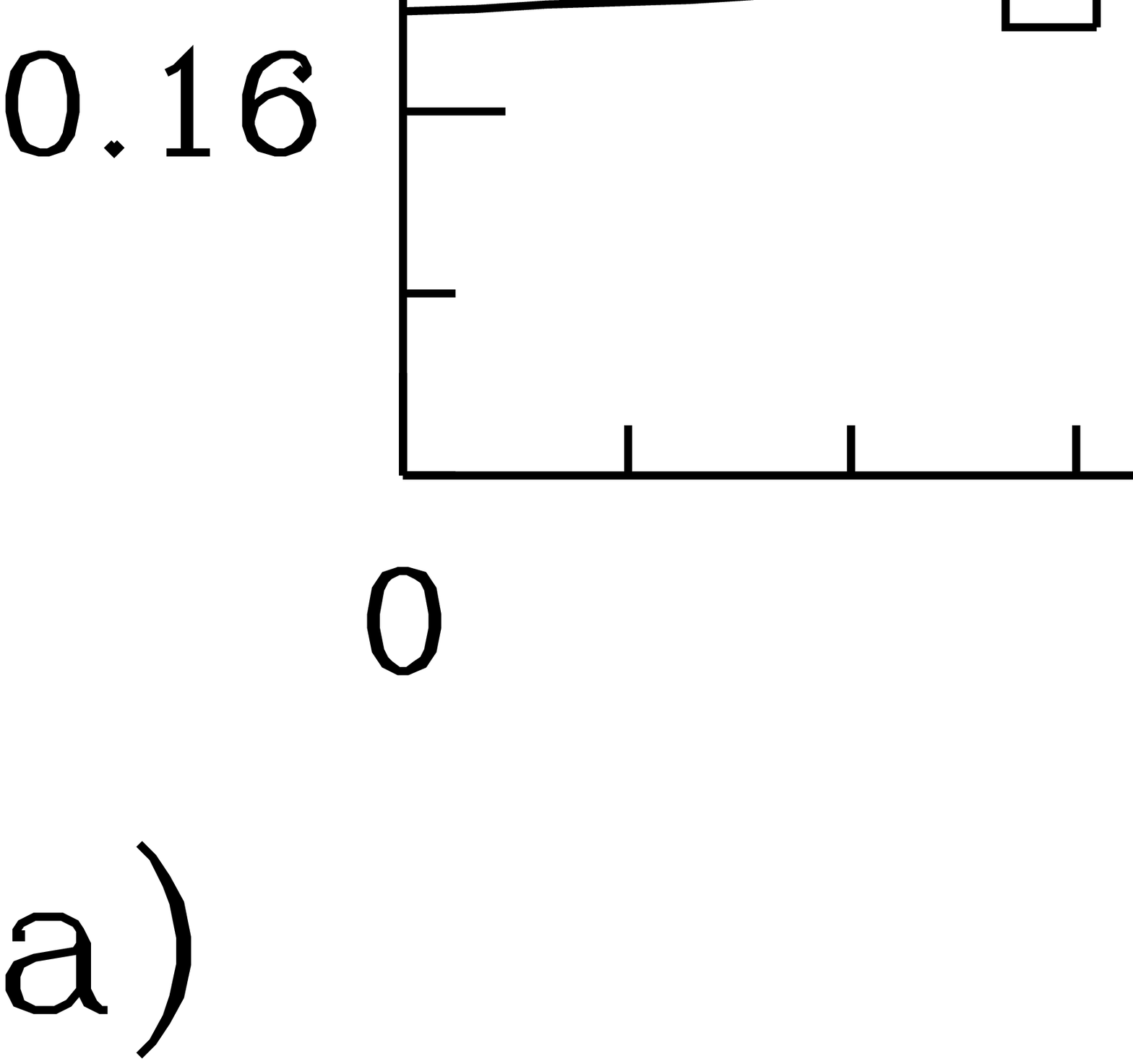} }
\vspace{0.7cm}
 \centerline{ \epsfysize=9.50cm \epsfbox{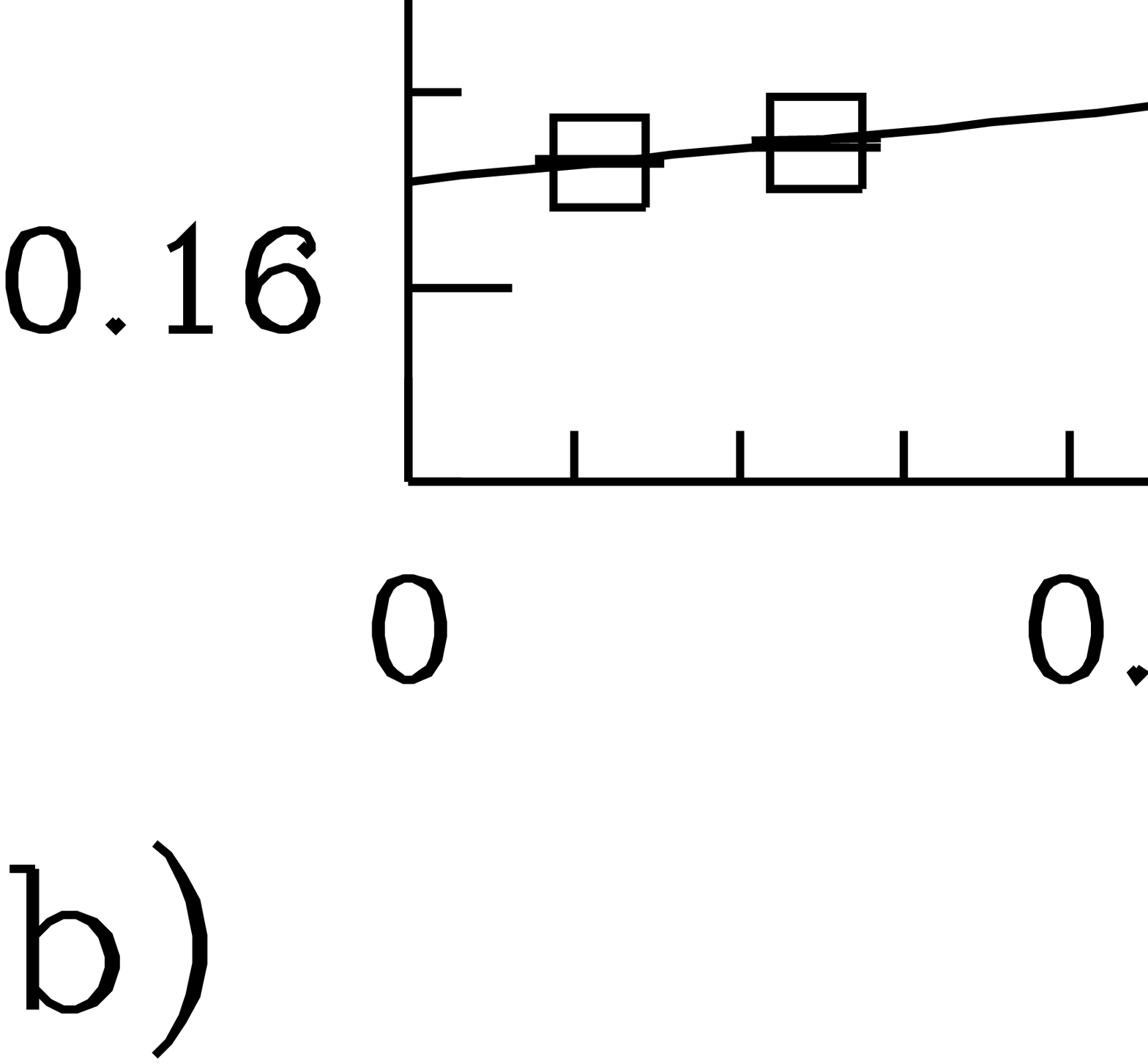} }
\end{figure}
\begin{figure} [ttt]
\vspace{-0.5cm}
 \centerline{ \epsfysize=9.50cm \epsfbox{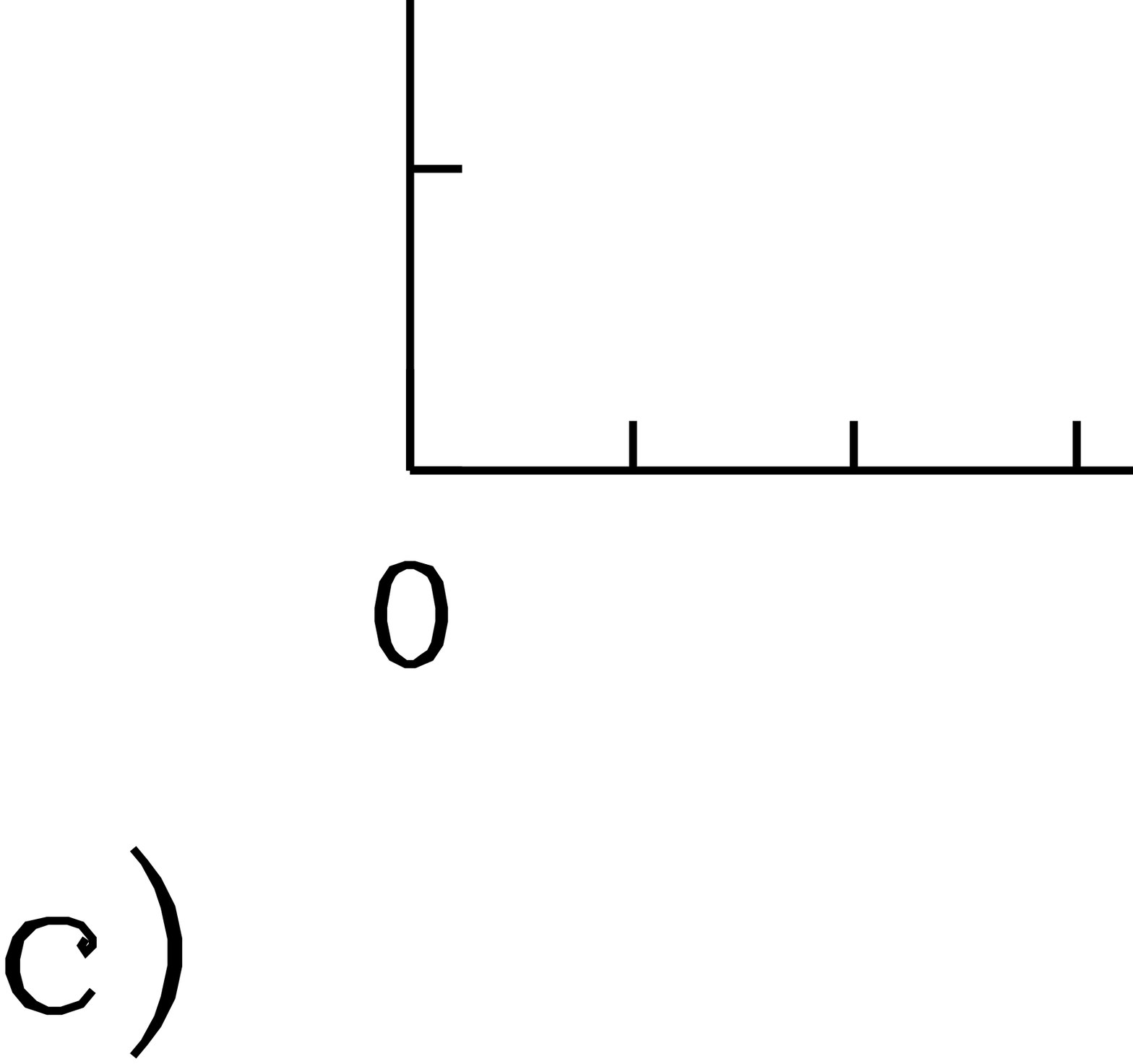} }
\vspace{-0.5cm}
\caption{{\em $S_{R^2}(\b/a) = \b \av{\sum_i R_i^2/N_2}_A$  as a function
              of $1/N_2$ for the bi-torus  (a), the
              sphere (b) and the torus (c). We have not
              included in these graphs the
               $S_{R^2}(\b/a)$ results on small volumes
              since they exhibit deviations from the straight line behavior.
              The symbols and lines in figs.~a-c correspond
              to  various values of $\b/a$:
              (a) $\Box : \b/a=0.61$,
              ${\rm o}: \b/a = 2.4$,
              $\triangle : \b/a=4.5$;
              (b) $\Box : \b/a=0.61$,
              $ \bullet: \b/a = 2.5$,
              $\triangle : \b/a=4.5$;
              (c) $\triangle : \b/a=2.4$. The full circles in fig. b have been
generated
              with the Hybrid Monte Carlo algorithm.
              The straight lines  have been obtained
              by fitting the numerical data to the ansatz
              ({\protect \eq{R2ANS}}).
              }}
\lb{fig3}
\end{figure}

In order to make contact with ref. \cc{GrHa91}, we have  used the scale
invariant measure, $\z=0$. Then relation (\ref{CAN}) is   not
satisfied, but we shall nevertheless define a string susceptibility from
the finite part of $\av{\sum_i R^2_i}_A$, i.e.
$\g_{{\rm str}}=c_1^{(0)}+2=c_1(\infty)+2$. We shall show below
that $c_1^{(0)}$ in fact appears not to change much when tuning $\z$ such that
(\eq{CAN}) is fulfilled.

To determine this susceptibility from our numerical data
we have to extrapolate the $c_1(\b/a)$ data to $\b/a =\infty$.
The results for $c_1(\b/a)$ are plotted in fig.~\ref{fig4}
as a function of $\b/a$ for the bi-torus (fig. a) and the sphere (fig. b).
The two plots show that it is difficult to extrapolate
reliably to large values of $\b/a$. For the
bi-torus we find from our data a lower limit
$\d\g=c_1^{(0)} \apgt 3.5$, provided that $c_1(\b/a)$
keeps monotonously increasing for $\b/a > 15$.
Using the heuristic fit ansatz
\be
c_1(x)=A+B/(x+C) \lb{ANS}
\ee
we find plausible values for $c_1^{(0)}$ in the range
$4.0 \lsim c_1^{(0)} \lsim  5.0$.  The curves obtained by fitting
the $c_1(\b/a)$ data
to the ansatz (\eq{ANS}) are represented in fig.~\ref{fig4} by the solid lines.

\begin{figure} [ttb]
\vspace{-1.2cm}
 \centerline{ \epsfysize= 9.0cm \epsfbox{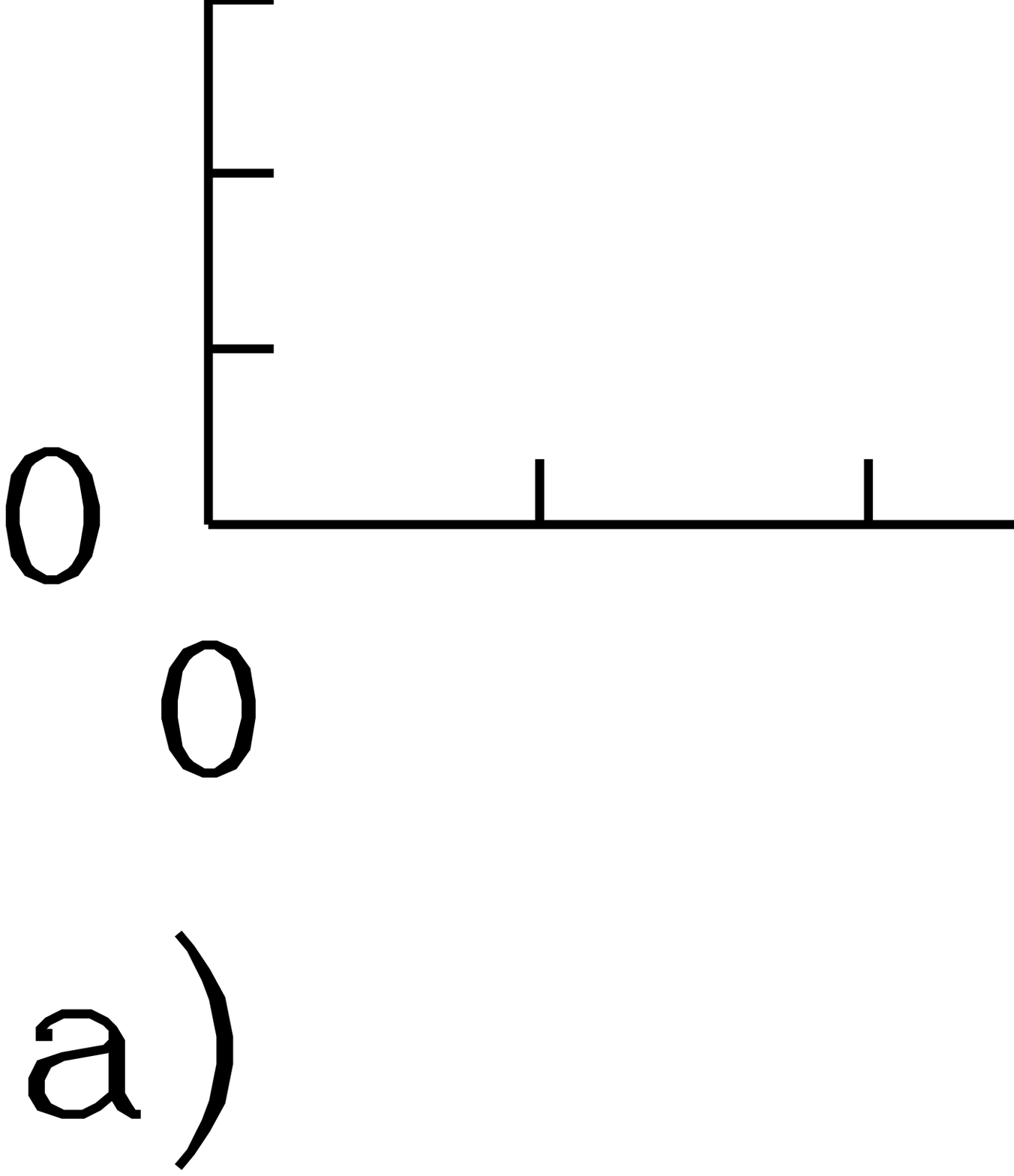} }
\vspace{0.5cm}
 \centerline{ \epsfysize= 9.0cm \epsfbox{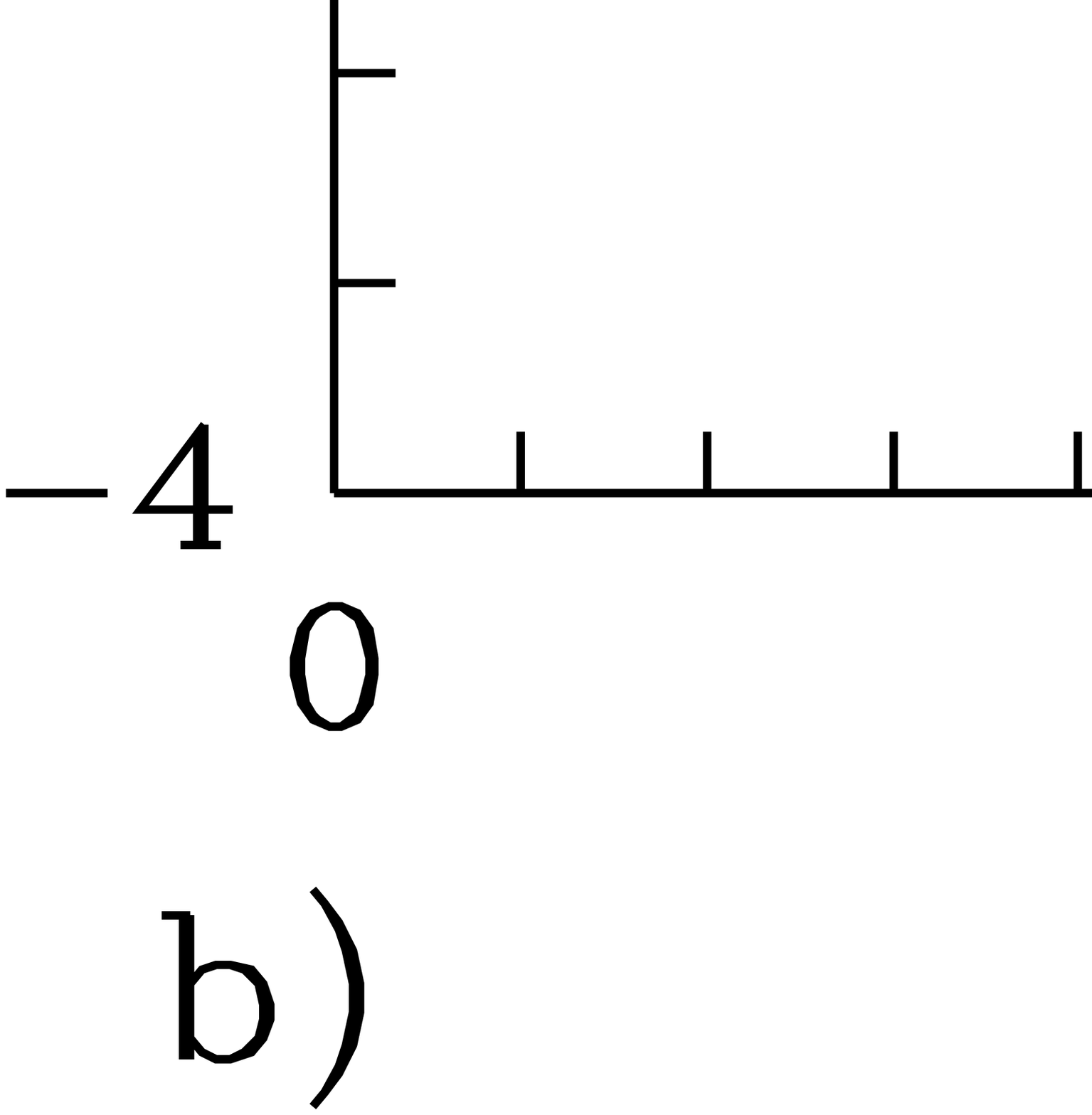} }
\vspace{-0.7cm}
\caption{{\em The coefficient $c_1(\b/a)$ as a function
              of $\b/a$ for the
              bi-torus (fig.~a) and the sphere (fig.~b).  Most of the numerical
              results have been obtained with  the
              histogramming method (open circles and triangle). The results
              represented
              in fig. b by the full circles have been obtained
              with the fixed area
              Hybrid Monte Carlo
              code.  The full lines are fits with the heuristic ansatz
              ({\protect \eq{ANS}}). The fit in fig.~a
              does not include the five points at $\b/a \approx 0.20$.
              The dashed lines
              represent the $\g_{\rm str}-2$ values in the continuum.
              The triangle in fig.~a  has been obtained
              at $\b/a=1.7$, $\z=-0.2$ which is closer to the point
              where ({\protect \eq{CAN}}) is fulfilled.
              }}
\lb{fig4}
\end{figure}

For the sphere we have computed $c_1(\b/a)$ only
for a few $\b/a$ values. Fig.~\ref{fig4}b shows
that the  $\b/a$ dependence of the coefficient $c_{1}(\b/a)$
is much weaker than for the bi-torus, but also
in this case $c_1(\b/a)$ appears to increase monotonously with $\b/a$.
The data which are represented
by the full circles have been obtained with the fixed area Hybrid Monte
Carlo code.
We have used the generalized triangle inequalities which
put a non-zero lower limit on the minimum triangle area ($\e = 0.01$), as
discussed in sect.~4.2.  As can be seen in fig.~\ref{fig4}b, such a small
value of $\e$ appears not to lead
to significant deviations compared with the Metropolis results.
{}From this figure we can read of
a lower bound for $c_1^{(0)}$ which is approximately equal to 3.5.

These results were obtained using the same scale invariant measure
as in ref. \cc{GrHa91}.
It is clear that the resulting values for $\g_{{\rm str}}=2+c_1^{(0)}$ are
in strong disagreement with the continuum
predictions for both topologies:
For the bi-torus we find $\g_{{\rm str}} \apgt 5.5$ whereas the continuum
value is
$\g_{{\rm str}}=4.5$ and for the sphere we find $\g_{{\rm str}} \apgt 5.5$
whereas the continuum value is $\g_{{\rm str}}=-0.5$.
The continuum values of $\g_{{\rm str}}-2$ are represented
in fig.~\ref{fig4}a and b by the horizontal dashed lines.
For the torus we find that
$c_1(\b/a) \approx 0$, independent of $\b/a$ and hence also $\d\g=c_1^{(0)}
\approx 0$, as in the continuum. This means that the partition
function obeys the scaling relation $Z(A) \propto A^{-1}$,
which, as we pointed out in sect.~2.1, also holds for the
classical model. Therefore it gives no information on the
contribution of quantum effects, and finding the correct scaling of the
Regge partition function for this case is of little significance.

The results for the bi-torus and sphere show that the claim of
ref. \cc{GrHa91}, that the Regge model with
a scale invariant measure reproduces the continuum values of the
string susceptibility, is not justified.
However, we have argued above that the value of $\z$ in the
measure has to be chosen such that the term $\propto N_1$ in the
expectation value of the action is canceled, cf. eq. (\ref{CAN}).
This is not the case when using the scale invariant measure:
e.g. we find for the bi-torus
$c_0(\b/a) =0.1072(4)$, $0.1627(7)$, $0.2096(2)$, $0.2225(2)$, $0.2316(2)$,
$0.2355(6)$
for $\b/a=0.192$, $0.61$, $2.4$, $4.5$, $8.6$, $12.6$ and clearly
(\eq{CAN}) is not fulfilled since $\z=0$ in our case.

To check if the numerical results change when adjusting $\z$ such that
$c_0+3\z/4$ becomes closer to zero, we have performed
a simulation at $\b=0.1$ and $\z=-0.2$ for the case of the bi-torus.
For $\b/a=1.7$ we find $c_0+3\z/4\approx 0.0904(4)$  which is indeed
almost by a factor two smaller than the value of $c_0+3\z/4$ obtained
for $\z=0$ at the same value of $\b/a$.
The result  for the coefficient $c_1$ is represented in fig.~\ref{fig4}a by the
triangle. Even though the error bar is rather large, it occurs that the
result falls on top of the fit curve to the $\z=0$ results which indicates that
the
coefficient $c_1(\b/a)$ does not depend on the measure parameter $\z$.
\section{Summary and final remarks}
In this paper we have investigated the Regge method of regularizing the
euclidean path integral for quantum gravity, using the two dimensional
model as a test case.
We have shown that a Regge skeleton can be viewed as a discretization
of a smooth continuum manifold, which is defined by a metric field
$g_{\m \n}$.
The  Regge action, including $R^2$ terms, approximates the
corresponding continuum action, up to discretization errors.
These discretization errors break diffeomorphism invariance, as
illustrated in eq. (\ref{JUNK}) for the cosmological term in the
action.

This suggests to separate the Regge degrees of freedom, the
edge lengths of the skeleton, into equivalence classes, as in the
continuum. We have given such a possible prescription in which skeletons are
equivalent if they are related by a ``gauge transformation''
$l\ra l^\x$, cf. eq. (\ref{REGGAU}). In a Gaussian approximation of the
action around flat space, these transformations are zero modes, and
correspond to the continuum gauge transformations \cc{RoWi81}.
The presense of these modes suggests that the measure in the Regge
path integral should include a gauge fixing and Faddeev-Popov term,
as in the continuum. However, it is not known how to implement this
in practical simulations, and so far only simple variations on a scale
invariant measure have been used.

Our numerical simulations have been done with such a simple measure.
One can derive scaling relations for the Regge path integral of
two dimensional $R^2$ gravity, which relates $d\log Z(A)/dA$ to the
expectation value of the curvature square term, evaluated at a fixed
value $A$ of the total area of the two dimensional surface \cc{GrHa91}.
We use two Monte Carlo methods to compute the  constraint
expectation values. In the first of these methods we simulate the unconstraint
path integral and implement the fixed area constraint a posteriori
by using a histogramming technique. The second method is
a Hybrid Monte Carlo algorithm with
which such a constraint model can be simulated directly.
This algorithm is very attractive, because it can easily be
vectorized and in comparison to the histogramming method it is more efficient
because it allows us to carry out a simulation at a chosen  value of $\b/a$.

{}From the exact scaling relations for the Regge partition function we
infer that the string susceptibility is determined from the part
of   $\b\av{R^2}_A + N_1\z/2$ that is finite after removing the
regularization, i.e. for $N_1\ra\infty$.
The first term is the expectation
value of the action at a fixed value of the area $A$ and $\z$ is
the measure parameter.
The identification of this  finite part is however ambiguous and depends
sensitively on the specific form of the measure used, as can be seen
by shifting the measure parameter $\z \ra \z+\z_1/N_1$. This is
particularly clear for $\b=0$, where we find the exact result
$\g_{\rm str}= \z_1/2$.

These considerations cast serious doubts on the claim made in ref. \cc{GrHa91}
that the Regge model with scale invariant measure reproduces the
continuum values of the string susceptibility. To settle this, we
have also performed a numerical simulation of this model: as
improvements over the earlier work, we
have done the simulation at a fixed value of the total area, we have
used a spherical topology with a more regular skeleton which only has
coordination numbers four and six, and we also included a topology with
genus $h=2$, the bi-torus. Otherwise we use the same scale invariant
measure and $R^2$ term as in ref. \cc{GrHa91}.
We have determined the coefficient $c_1(\b/a)$ of the $1/N_2$ term in
the average action to a high accuracy for all three topologies and for
several values of the  coupling constant $\b/a$.

In ref. \cc{GrHa91} the coefficient $c_1$ was wrongly     identified
with the string susceptibility.
It turns out that $c_1$
depends strongly on $\b/a$ for the case of the bi-torus.
For the sphere the $\b/a$ dependence is much weaker
(see fig.~\ref{fig4}) and for the torus
we find no $\b/a$ dependence. We have argued that
the determination of the string susceptibility requires that $\b/a\ra \infty$
and that $\z$ is tuned such that eq. (\ref{CAN}) is satisfied.
In our simulations we have followed ref. \cc{GrHa91} and chosen the
scale invariant measure $\z=0$, for which eq. (\ref{CAN}) is not
fulfilled.
Simulations at a negative value of $\z$ at which this
equality is almost satisfied indicate that $c_1$ is not affected by a
small change in $\z$. Assuming that we do not include a term $\z_1/N_1$
in the measure, we can then from the $c_1(\b/a)$ values shown in
fig.~\ref{fig4} infer lower bounds on  $\g_{\rm str}$. Both for the
sphere and the bi-torus we find
$\g_{{\rm str}}\apgt 5.5$ which is substantially
larger than the continuum values given in eq.~(\eq{STRING}).
Only for the torus the string susceptibility reproduces the continuum
value, but here the naive scaling holds, which does not involve
quantum fluctuations.

A likely explanation for this failure of the Regge approach is
the contribution of the gauge degrees of freedom, the $\x$'s of eq.
(\ref{REGGAU}). These modes do not decouple form the action, because the
Regge action is not gauge invariant
(this is only approximately the case for small transformations of a
smooth skeleton \cc{RoWi81}). The gauge modes have also
not been removed by including a
gauge fixing and Faddeev-Popov term in the integration measure.
If this is indeed the source of the failure of the Regge approach in
two dimensional gravity, one would expect similar problems also in
four dimensions.

To test this hypothesis and to improve the Regge approach, one could try
to discretize the continuum
model, including these gauge fixing and Faddeev-Popov terms.
Such an approach faces many complications: The main obstacle is
that one would have to find a practical
method to simulate  an action that includes such
a gauge fixing term and Faddeev-Popov ghosts.
Another serious complication arises from    the breaking of the gauge
(diffeomorphism) invariance which  will produce gauge variant terms under
renormalization. These terms have to be canceled by adding suitable
counterterms.
Furthermore, such gauge fixing and
Faddeev-Popov ghost action  expressed in terms of the Regge edge
lengths, will be very cumbersome. Then it might be easier to  abandon the
Regge approach and
use a more straightforward regularization of the path integral by
discretizing the metric field $g_{\m \n}$ on a hypercubic lattice.
\subsubsection*{Acknowledgement}
We have benefited from discussions with J.~Kuti and C.~Liu and thank
them for many useful suggestions. Further we also like to thank W.~Beirl,
B.~Berg, M.~Golterman, U.~Heller, J.~Smit and M.~Visser for useful discussions.
This work was supported by the DOE under grant DE-FG03-91ER40546
and by the TNLRC under grant RGFY93-206. Some of the numerical
simulations were done at the Livermore  National Laboratory with
DOE support for supercomputer resources and on the CRAY C90 at the San
Diego Supercomputer Center.

\end{document}